\documentclass[twocolumn]{aastex631}
\usepackage{lineno}
\shorttitle{Galaxy populations in groups and clusters}
\shortauthors{J. Meng et al.}

\begin{document}

\title{Galaxy populations in 
groups and clusters: evidence for a characteristic 
stellar mass scale at $M_\ast\sim 10^{9.5}M_\odot$}

\correspondingauthor{Jiacheng Meng \& Cheng Li}
\email{mengjc18@mails.tsinghua.edu.cn}
\email{cli2015@tsinghua.edu.cn}

\author[0000-0002-1974-905X]{Jiacheng Meng}
\affiliation{Department of Astronomy, Tsinghua University, Beijing 100084, China}

\author[0000-0002-8711-8970]{Cheng Li}
\affiliation{Department of Astronomy, Tsinghua University, Beijing 100084, China}


\author[0000-0001-5356-2419]{H.J. Mo}
\affiliation{Department of Astronomy, University of Massachusetts Amherst, MA 01003, USA}

\author[0000-0002-4597-5798]{Yangyao Chen}
\affiliation{Department of Astronomy, University of Science and Technology of China, Hefei Anhui 230026, China}

\author[0000-0003-2405-5930]{Zhen Jiang}
\affiliation{Department of Astronomy, Tsinghua University, Beijing 100084, China}

\author[0000-0003-3864-068X]{Lizhi Xie}
\affiliation{Tianjin Normal University, Binshuixidao 393, 300387, Tianjin, China}
\affiliation{INAF - Astronomical Observatory of Trieste, via G.B. Tiepolo 11, I- 34143 Trieste, Italy}



\begin{abstract}

We use the most recent data release (DR9) of the DESI legacy imaging survey and SDSS galaxy groups to measure the conditional 
  luminosity function (CLF) for groups with halo mass $M_{\rm h}\ge 10^{12}M_{\odot}$
  and redshift $0.01\le z\le 0.08$, down to a limiting $r$-band magnitude of 
  $M_{\rm r}=-10\sim-12$. 
  For a given halo mass we measure the CLF for the 
  total satellite population, as well as separately 
  for the red and blue populations classified using the $(g-z)$ color. We find a clear 
  faint-end upturn in the CLF of red satellites,
  with a slope $\alpha\approx-1.8$ which is almost independent of halo mass. 
  This faint-end upturn is not seen for blue satellites and for the total population. Our stellar population synthesis modeling shows that the $(g-z)$ color 
  provides a clean red/blue division, and that 
  group galaxies in the red population defined by $(g-z)$ are all dominated by old 
  stellar populations. The fraction of old galaxies as a function 
  of galaxy luminosity shows a minimum at a luminosity 
  $M_{\rm r}\sim-18$, corresponding to a stellar 
  mass $M_\ast\sim10^{9.5}M_\odot$. 
  This mass scale is independent of halo mass and is
  comparable to the characteristic luminosity at which galaxies show a dichotomy in surface brightness and size, suggesting that the dichotomy in the old 
  fraction and in galaxy structure may 
  have a common origin. 
  The rising of the old fraction at the faint end for Milky Way (MW)-sized halos found here is in good agreement with the quenched fraction measured both for the 
  MW/M31 system and from the ELVES survey.
  We discuss the implications of our results for 
  the formation and evolution of low-mass galaxies, 
  and for the stellar mass functions of low-mass 
  galaxies to be observed at high redshift. 

\end{abstract}



\section{Introduction} \label{sec:intro}
\defcitealias{Lan2016}{L16}

In the current paradigm of structure and galaxy formation, 
galaxies form in dark matter halos through a two-stage process, in which 
dark halos form in the first place by gravitational instability of initial density 
perturbations, followed by the formation of galaxies at the centers of 
dark halos through gas cooling and condensation 
\citep[][and references therein]{White1978,MoBoschWhite2010}.
Halos grow in dark matter mass by both mass accretion 
and halo-halo mergers. A galaxy at the center of its host halo,  
called the central galaxy, grows its stellar mass by forming stars from 
the cooled gas and by accreting stars that have formed elsewhere.  
At a later time, a halo may fall into a bigger halo and become a subhalo, 
and the central galaxy hosted by it then becomes a satellite galaxy in  
the new host. As a satellite, it is affected by environmental effects 
in the host halo, such as tidal stripping and ram-pressure stripping, 
which may effectively reduce the hot and even cold gas contents of the 
galaxy, shutting down its star formation and making it quenched and red. 
On the other hand, physical processes internal to the galaxy, 
such as energy feedback from supernovae and active galactic nuclei may 
be able to heat and/or eject the cold gas, which may also quench the 
star formation and make the galaxy red. Thus, the formation and evolution of 
galaxies are expected to be driven by a variety of factors, including the assembly 
history and properties of (sub)halos of galaxies and internal processes
and properties of galaxies. In order to obtain a full understanding of galaxy 
formation and evolution, therefore, it is crucial to establish a statistical link 
of the properties of galaxies to their dark matter (sub)halos. 

To this end, large amounts of effort have been devoted to measuring the luminosity 
function (LF) and stellar mass function (SMF) of galaxies  
in clusters and groups of galaxies, so as to quantify the galaxy population 
in halos of different mass and in systems of different richness. 
Early studies were usually limited to nearby rich clusters, and the  
LFs of cluster galaxies were found to be roughly consistent with a Schechter form
in the bright end, but with some sign of a steepening in the faint-end  
\citep[e.g.][]{Binggeli1988,Driver1994,Bernstein1995,dePropris1995,Yagi2002,Parolin2003,
	Popesso2005,Popesso2006,Barkhouse2007,Barkhouse2009}. 
This faint-end upturn was also found in galaxy groups in the local Universe
\citep[e.g.][]{Zandivarez2006,Yang2009CLF,Robotham2010,Zandivarez2011},
mainly from the spectroscopic galaxy sample of the Sloan Digital Sky Survey 
\citep[SDSS;][]{York2000}. However, these findings of a significant 
faint-end upturn were questioned by other investigations using 
data from, e.g., the SDSS~\citep{,Hansen2009,deFilippis2011} and the
HST~\citep{Harsono2009}. Results obtained for individual systems in the nearby Universe 
also show significant variance \citep[see][for a review]{Boselli2014}, 
with a significant faint-end upturn present for some clusters, 
such as A1689 \citep{Banados2010}, Coma \citep{Mobasher2003,Yamanoi2012} and
Abell 119 \citep{Lee2016}), but absent for some others,  
such as Virgo \citep{Rines2008,Lieder2012,Ferrarese2016}, 
Hydra \citep{Yamanoi2007,Misgeld2008,Misgeld2009} and 
Abell 85 \citep{Agulli2014}. More recently, \citet{Hashimoto2022} 
measured the LFs of dwarf galaxies in 33 galaxy clusters at $z\sim0.15-0.3$
using data obtained from the Subaru Superime-Cam imaging survey, and 
found that the faint-end slope is rather flat, with $\alpha=-1.2\sim-1.4$.
In contrast, analyses using central galaxies and groups selected
from the SDSS spectroscopic data in combination with satellite galaxies 
sampled by photometric data from SDSS \citep{Lan2016}, from DESI \citep{Tinker2021} 
and from a combination of SDSS, DESI and HSC \citep{Wang2021satMW}
revealed that the faint end of the conditional luminosity 
function (CLF) of galaxies shows a significant upturn for halos of 
all mass, and the upturn is particularly strong for red satellites, 
with $\alpha\approx -1.8$ \citep{Lan2016}.

 These measurements provide important information about the galaxy-halo 
connection and on the physical processes behind it
(e.g. see \citealt{WechslerTinker2018} for a review). In particular,
the faint-end slope of the LFs in groups/clusters provides a unique way to 
understand the formation and evolution of low-mass galaxies. 
For instance, in the empirical model developed by \citet{Lu2014empirical} 
and \citet{Lu2015empirical}, a high efficiency of star formation at $z>2$ 
is required for their model to reproduce the faint-end upturn in the 
cluster galaxy LF obtained by \citet{Popesso2006}. In \citet{Lan2016}, 
the subhalo abundance matching method was used to relate the faint-end slope 
($\alpha$) of the CLFs of the SDSS galaxy groups to the power-law slope 
($\beta$) of the galaxy luminosity-subhalo mass relation ($L\propto m^\beta$), 
They obtained $L\propto m$ for low-mass red satellites in comparison to $L\propto m^{3/2}$ 
for their blue counterparts. The relation $L\propto m$ indicates 
that low-mass red satellites formed their stars with an efficiency 
independent of halo mass, a result that has important implications for star 
formation and feedback in these systems. The observed results are 
consistent with the empirical model of \citet{Lu2014empirical} and \citet{Lu2015empirical}
and the pre-heating model of \citet{MoMao2002} and \citet{MoMao2004}, 
both predicting enhanced star formation efficiency in low-mass halos at high 
redshift. Clearly, reliable observational results are required to test these predictions. 
In addition, the satellite population in clusters of galaxies 
may also be related to the origin of the chemical enrichment of the intracluster 
medium (ICM). \citet{Trentham1994} suggested that the precursors of dwarf galaxies,  
in which supernova-driven winds may expel gas effectively, can account for 
the enrichment of the ICM if the faint-end slope is steep, $\alpha\sim -1.9$. 
However, \citet{Gibson1997} showed that the precursors of the dwarf galaxies cannot
account for the chemical content of the ICM even if the faint-end 
slope is steep. Thus, the importance of the satellite population for  
the enrichment of the ICM is still uncertain.

In this paper, we attempt to obtain reliable measurements of the CLF down 
to unprecedented limits for galaxy groups in the local Universe. 
The CLF was originally proposed by \citet{Yang2003clf} and has been 
studied extensively using both spectroscopic and photometric samples 
\citep[e.g.][]{vandenBosch2003,Hansen2005,Cooray2006,vandenBosch2007,Hansen2009,Yang2009CLF,
	Guo2011CLF,Wang2012,Sales2013,Lan2016,Guo2018,Vazquez-Mata2020,
	Tinker2021,Wang2021satMW}.
We use central galaxies identified by applying the halo-based group finder 
of \citet{Yang2005} to the galaxy sample of SDSS data release 7 
\citep{Yang2007}. CLFs have also been measured for SDSS groups by 
\citet[][hereafter L16]{Lan2016} using the SDSS photometric sample to 
trace the satellite population, and by \citet{Tinker2021} 
using an earlier data release of the DESI imaging survey. Here, we use the 
most recent imaging data from the DESI legacy survey 
\citep[DESI-DR9;][]{DESIlegacy2019}, which is deeper and provides better photometry 
than data used in previous studies. In addition, we will use the $(g-z)$ color 
instead of the commonly-used $(u-r)$ color to separate galaxies into red and blue 
populations. As we will show, the $(u-r)$ is sensitive to young stellar populations, 
and it may mis-classify ``old'' galaxies dominated by old stellar populations 
as blue galaxies due to the contamination of a small fraction of young populations, 
particularly for low-mass galaxies of low metallicity. The $(g-z)$ provides
a cleaner color division, which allows us to measure the CLFs of faint galaxies 
more reliably for both the old and young satellite populations. Our analysis 
leads to the finding of a characteristic mass scale at 
$M_\ast\sim10^{9.5}M_\odot$ in the relation between the old fraction and 
the stellar mass, and that the mass scale is quite independent of halo mass. 
This indicates a dichotomy of the satellite population in quenching processes. 

The paper is organized as followed. In \autoref{sec:data} we describe the 
imaging data and the group catalogue used in our analyses. 
\autoref{sec:member} presents our measurements of the conditional 
luminosity functions. In \autoref{sec:old_population} we study 
the old satellite population versus their young counterpart,  
and use stellar population synthesis models to explore the origin 
of the color bimodality observed for galaxies in groups.  
We discuss our results in \autoref{sec:discussion} and 
summarize them in \autoref{sec:summary}.
Throughout the paper, we assume a $\Lambda$CDM cosmology with 
$\Omega_{\rm m}=0.275$ and $H_{0}=70.2{\rm km s^{-1} Mpc^{-1}}$ 
following the WMAP7 results \citep[][]{Komatsu2011WMAP7}. We define the mass 
and radius of a dark matter halo so that the mean mass density within the 
radius is 200 times the mean density of the universe. 
For convenience, we use $M_{\rm h}$ and $r_{\rm h}$, instead of $M_{\rm 200m}$ 
and $r_{\rm 200m}$, to denote these quantities. 

\section{Data}
\label{sec:data}

\subsection{SDSS galaxies and groups}
\label{sec:sdss}

We use galaxy groups identified by \citet{Yang2007} from the NYU value-added 
galaxy catalogue \citet[NYU-VAGC;][]{Blanton2005a}, a catalog of low-$z$ galaxies 
based on SDSS data release 7 \citep[DR7;][]{SDSSDR72009}. 
These groups were selected using 
the halo-based group finder of \citet{Yang2005}, which uses an iterative
method adapted to the properties of dark matter halos. 
Unlike the Friend-of-Friend (FoF) method, the halo-based finder is capable 
of identifying poor groups, even those with only one member. We use the group 
catalogue constructed from galaxy Sample II, which covers an area of 7748 
$\rm{deg^2}$ and contains galaxies from the SDSS main sample with redshift 
$0.01\le z\le 0.2$ and with redshift completeness $\mathcal{C}>0.7$. 
Sample II also includes 7091 galaxies with redshift from other surveys, 
such as the 2dFGRS \citep[][]{Colless20012dF}, but it does not include 
galaxies missing  redshift measurements due to fiber collision. 
\citet{Yang2007} used two methods to assign halo mass to galaxy groups,
based on the ranks of either the total stellar mass or the total
$r$-band luminosity of all group members with 
$M_{\rm r}-5\log h\le 19.5$. Here we use halo masses, $M_{\rm h}$, 
estimated from the total stellar mass. The halo sample with $M_{\rm h}\ge 10^{12}M_{\odot}$ 
are complete at $z\le 0.08$. We thus use the group/halo sample 
with $0.01\le z\le 0.08$ and $M_{\rm h}\le 10^{12}M_{\odot}$ for  our analyses,
unless otherwise stated.

In our analyses, we also use both spectroscopic and photometric galaxies 
from the SDSS. Spectroscopic galaxies are used to estimate {\it K}-corrections 
for photometric galaxies and to check results obtained from photometric samples.  
The spectroscopic galaxies used are also from NYU-VAGC, and
{\it K}-corrections of galaxies are obtained using
$\mathtt{kcorrect\enspace v4\_1\_4}$ \citep[][]{Blanton2007kcorrect}.
As we will describe below, the DESI imaging data, which we will also use for 
our analyses, provides only photometry in the $g$, $r$ and $z$ bands. We will supplement 
the data with the $u$-band photometry from the SDSS when needed. 
The SDSS photometric galaxies are selected from the SDSS data release 16 \citep[DR16;][]{SDSSDR162020}. 
We use galaxies with $r$-band model magnitudes $r<21$ to guarantee 
completeness of the data. 
We note that SDSS photometry is used only for the $u$-band; magnitudes in all other 
bands are from the DESI image data.

We use two methods to do the {\it K}-correction for photometric galaxies, both based on 
the {\it K}-correction values of the SDSS main sample provided in the
NYU-VAGC. In the first method, we create a grid in observed $(g-r)$ color 
versus $(r-z)$ color space to obtain the median {\it K}-correction in each band 
as a function of redshift for each of the color-color bins. The value of the 
{\it K}-correction for each photometric galaxy in a color-color bin is then estimated 
using the redshift of the galaxy group to which the galaxy is assigned. 
The second method is to use the nearest galaxies in the SDSS 
spectroscopic sample. For each photometric galaxy, we find its nearest neighbor,  
in the spectroscopic sample, which is closest to it in the observed 
$(g-r)$-$(r-z)$-redshift space, and use the {\it K}-correction value of the nearest 
spectroscopic neighbor as its {\it K}-correction. Here the `distance' (which decides 
the nearest neighbor of a galaxy) is defined as the square root of  
$\Delta^2\left(g-r\right)+\Delta^2\left(r-z\right)+\Delta^2 z$. 
The redshift used for the photometric galaxies are the redshift of the galaxy group. 
We find that the two methods give very similar results, and 
our following presentation is based on the second method. 

\subsection{The DESI imaging data}
\label{sec:desi}

The DESI legacy imaging survey \citep[DESI-DR9,][]{DESIlegacy2019} 
consists of three parts: the Beijing-Arizona Sky Survey (BASS), Mayall $z$-band Legacy 
Survey (MzLS), and DECam Legacy Survey (DECaLS). The depths at 
$5\sigma$ are about $g=24.7$, $r=23.9$, and $z=23.0$, respectively,   
and the total sky coverage is about $14,000 {\rm deg^2}$. We use galaxies
that have $r<23$ and meet the following selection criteria:
\begin{enumerate}
    \item {\tt TYPE} = ``REX'',``EXP'', ``DEV'' or ``SER'';
    \item {\tt FRACFLUX\_R} $<2$;
    \item {\tt FRACMASKED\_R} $<0.6$.
\end{enumerate}
Here {\tt TYPE} represents the morphological type of the sources, and the four
types listed above are considered as galaxies. {\tt FRACFLUX\_R} represents 
the profile-weighted fraction of the $r$-band flux of a source that comes from 
other sources. We have tested using other criteria, such as 
{\tt FRACFLUX\_R} $<2.5$ or $3$, and found no significant impact on our results. 
{\tt FRACMASKED\_R} represents 
the fraction of masked pixels of an object in the $r$-band, and here we follow the 
criterion used in \citet{Tinker2021}. 
We find that some sources listed in the catalogue are substructures of 
bigger galaxies. We use the following method to remove them. For each galaxy in 
DESI-DR9, we use its half light radius, {\tt SHAPE\_R}, and its orientation
and axis ratio specified by {\tt SHAPE\_E1} and {\tt SHAPE\_E2} 
to determine an ellipse with a linear size twice as large as that 
given by the half-light radius. We exclude all galaxies that reside in the 
ellipse of another galaxy. The factor of two in the linear size of the ellipse 
is chosen somewhat arbitrarily, but our tests using other factors 
show that our results are not sensitive to the choice. 
DESI-DR9 also provides random samples of the survey and we use them 
to model the survey mask.

For some of our analyses, we also need the $u$-band photometry, which is not provided by DESI-DR9. 
As a remedy, we use the $u$-band photometry from SDSS to supplement the data. 
Galaxies are cross-matched using centroid positions, (RA, Dec), provided 
by the two catalogs. Because the SDSS survey is shallower than DESI-DR9, 
we only match galaxies with $r<21$ to guarantee the completeness of 
the SDSS data. Thus, only galaxies with $r<21$ are used in the analyses
that involve the $u$-band data.

Since no redshift is available for photometric galaxies, 
{\it K}-corrections cannot be applied directly for them. As 
mentioned above, we use the {\it K}-corrections obtained from the 
spectroscopic sample and matching in color-color space to estimate the 
{\it K}-corrections of photometric galaxies. 
There are some systematic offsets between the SDSS and DESI magnitudes. 
To account for these offsets,  we first match SDSS spectroscopic galaxies 
to the DESI image data to get their DESI-DR9 $g$, $r$ and $z$ magnitudes, 
and assume that galaxies at the same position in $(g-r)$-$(r-z)$-redshift 
space have a similar {\it K}-correction. For each photometric galaxy, which 
is determined to be associated to a given group (see below), 
we then identify its nearest spectroscopic galaxy in $(g-r)$-$(r-z)$-redshift 
space, where the redshift of the photometric galaxy is that of the group.
Finally, we use the {\it K}-correction of the spectroscopic galaxy 
as the {\it K}-correction for the photometric galaxy in question.  

\section{Galaxies in halos of galaxy groups} \label{sec:member}

\subsection{Method to identify member galaxies}
\label{sec:method}

Because there is no spectroscopic redshift for photometric galaxies, it is difficult 
to identify member galaxies in a galaxy group in a deterministic way. 
Here we use the same method as used in \citetalias{Lan2016} to identify 
member galaxies statistically. Our goal is to obtain the  distribution 
of member galaxies in the space spanned by galaxy properties, represented 
collectively by $\vec{\mathbf{q}}$, in groups of given halo mass, i.e. 
the conditional distribution function, $\Phi\left(\vec{\mathbf{q}}\mid M_{\rm h}\right)$. 
To this end, we cross-correlate galaxy groups with photometric galaxies. 
Since galaxy groups are identified from the spectroscopic data, their redshifts 
are known. For the $i$-th group with redshift $z_{\rm i}$, we identify all 
photometric galaxies within the angular radius of the group (obtained
from its physical radius and redshift), and assign the redshift $z_{\rm i}$
to all of them. The redshift is then used to calculate 
physical quantities of the photometric galaxy, such as absolute 
magnitudes and rest-frame colors. Finally, we obtain the conditional
function, $\Phi_{\rm grp,i}\left(\vec{\mathbf{q}}\mid M_{\rm h}\right)$, 
of the property $\vec{\mathbf{q}}$ of interest for group $i$ from the corresponding 
observational quantity and $z_{\rm i}$. 
Since we count all galaxies within the projected halo radius $r_{\rm h}$, 
the distribution function obtained above contains galaxies 
that are not true members, but foreground/background galaxies that are selected 
because of projection. For brevity, we will refer to these foreground/background 
galaxies collectively as background galaxies. To subtract the contribution of 
the background galaxies, we use an annulus with inner and outer radii 
of $2.5r_{\rm h}$ and $3r_{\rm h}$ around the group in question, and identify 
all photometric galaxies, all assumed to be at $z_{\rm i}$, within the annulus. 
Here again, we use the group redshift, $z_{\rm i}$ to estimate $\vec{\mathbf{q}}$
from observational quantities. The corresponding galaxy distribution, 
$\Phi_{\rm bkg,i}\left(\vec{\mathbf{q}}\mid M_{\rm h}\right)$, can then be estimated.  
Averaging over all groups with similar halo mass, we get the conditional 
distribution function, with the background subtracted:  
\begin{equation}
    \Phi\left(\vec{\mathbf{q}}\mid M_{\rm h}\right)=
    \left \langle \Phi_{\rm grp,i}\left(\vec{\mathbf{q}}\mid M_{\rm h}\right)
    -f_{\rm A,i}\times \Phi_{\rm bkg,i}\left(\vec{\mathbf{q}}\mid M_{\rm h}\right) \right \rangle_{\rm i},
\end{equation}
where $f_{\rm A,i}$ is the area ratio between the group and the annulus.

\citetalias{Lan2016} used two methods to estimate the background contribution. 
The first is a local estimator, which uses galaxies in annulus  
around individual groups, as described above. The second one is a  
global estimator, which uses the average of the galaxy counts 
in eight randomly selected circles, each with a radius corresponding to the 
halo radius $r_{\rm h}$, to estimate the background for the galaxy group in question.  
We have made tests using the two methods and found that the 
local estimator performs better. The global estimator 
ignores background fluctuations from place to place, leading to an 
underestimate of the conditional distribution functions. 
Because of this, we adopt the local estimator for our analyses.

In our analysis, we use 200 bootstrap re-samplings 
of groups to estimate measurement errors. Previous investigations have shown 
that clustering of background galaxies may affect error estimates
\citep[][]{Huang1997,Driver2003,Pracy2004}. \citet{Driver2003} 
proposed a formula to model the effects of clustering of background galaxies. 
In our analysis, an annulus outside the group region is used 
as the reference region for background subtraction. 
Thus, for each cluster/group of an angular size, the number count 
within its aperture consists of two parts: that of its member
galaxies, $N_C$, and that of background galaxies $N_B$, 
with the latter estimated from the count of galaxies in the annulus, 
$N_{\rm A}$. To estimate the impact of clustering on our 
background subtraction, we first randomly select positions across 
the DESI sky coverage. We then obtain the number counts of galaxies, 
$N_B$, within an aperture of radius $\theta$ around these positions, 
and the corresponding counts, $N_A$, in the annulus between 
$2.5\theta$ and $3\theta$. We estimate the average and variance of 
$\left(N_{B}-f_{A}N_{A}\right)$ as functions of both $\theta$ and the apparent 
magnitude $m_{\rm r}$. We find that the averages are all around 0, 
suggesting that our method to subtract the background is unbiased. 
As a test, we include the variance of $\left(N_{B}-f_{A}N_{A}\right)$ 
as an error in the background subtraction of each group according to 
its angular size and redshift (to link $M_r$ and $m_r$) in the 
error estimate of the CLF. The test result shows that the errors obtained 
by including this variance are comparable to the bootstrap errors, 
indicating that our error estimate is able to deal with 
effects of galaxy clustering in the background. 

\subsection{Conditional luminosity functions}
\label{sec:clf}

\begin{figure*}
    \centering
	\includegraphics[width=0.8\textwidth]{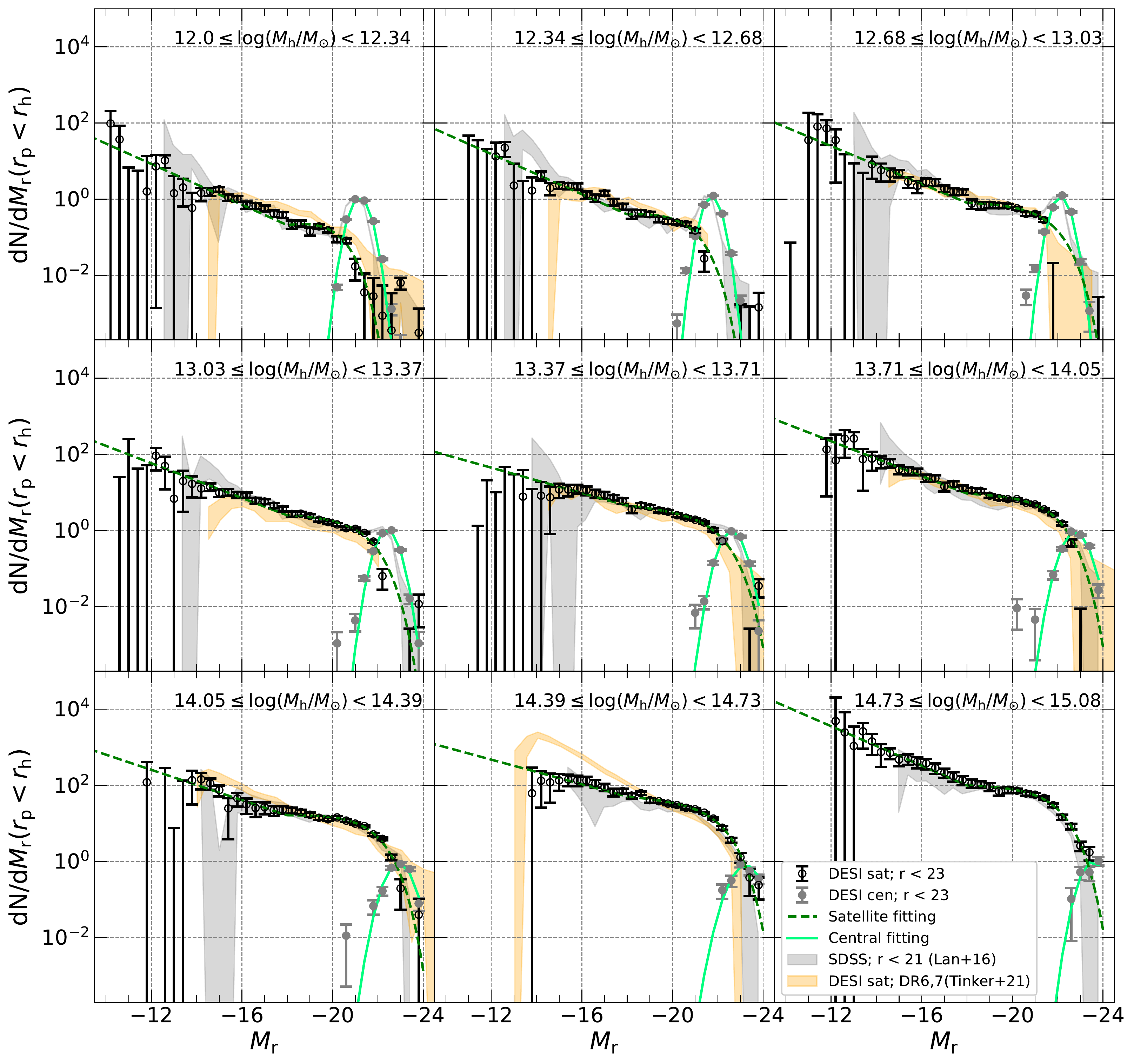}
    \caption{The grey and black dots represent our measurements of the luminosity functions for central 
    galaxies and satellite galaxies in galaxy groups at $0.01 \le z \le 0.08$ with the DESI imaging data. 
    The grey shaded regions represent the results of total luminosity functions in galaxy groups 
    in the redshift range $0.01 < z < 0.05$ obtained by \citet{Lan2016} using SDSS imaging data. 
    The orange shaded regions represent the results of the luminosity function of satellite galaxies 
    obtained from \citet{Tinker2021} using the DESI-DR6,7 imaging data. The green solid and dash lines are the 
    fitting results for our central and satellite luminosity functions respectively.}
    \label{fig:clf}
\end{figure*}

\begin{figure*}
    \centering
	\includegraphics[width=0.8\textwidth]{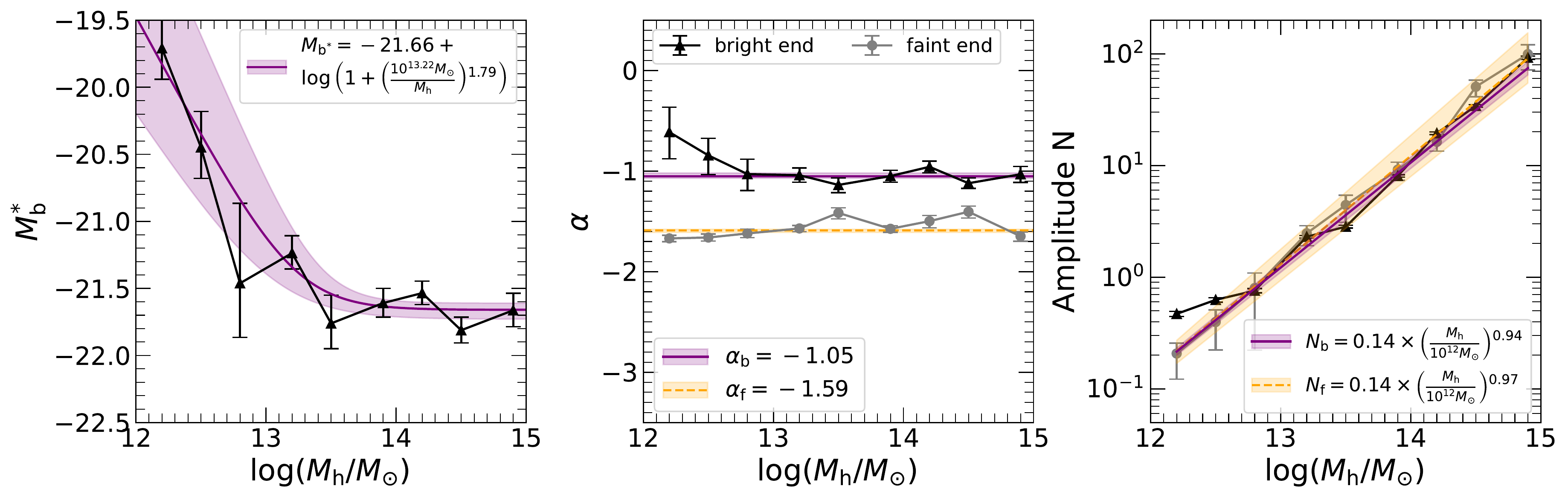}
    \caption{The figures show the model parameters for fitting the conditional luminosity functions of the whole sample as the function of halo mass. The errorbars represent the 1-$\sigma$ scatters of the parameters. The black and grey dots show the results for the bright end and faint end. We choose the proper function forms to describe the dependence of the parameters on the halo mass. The functions are written in the figures shown as the  purple and orange lines for the bright and faint end respectively.}
    \label{fig:para_whole}
\end{figure*}

We estimate the conditional luminosity functions (CLF) with the method described 
in \autoref{sec:method} using the DESI-DR9 imaging data and the SDSS galaxy 
group catalog described in \autoref{sec:data}. 
In this case, the galaxy property vector $\vec{\mathbf{q}}$ is the {\it K}-corrected
 $r$-band absolute magnitude: $\vec{\mathbf{q}}=\left\{M_{\rm r}\right\}$. 
To minimize observational selection effects, we use a group sample defined as follows. 
First, any group that has more than 5\% of its sky projection outside the DESI-DR9 
sky coverage is discarded. This step is achieved by using the random sample 
provided by the DESI-DR9. Second, we mask out a region around the SDSS Great Wall. 
The reason for this is that the distribution of galaxy groups in this region is 
crowded and some of the groups overlap with each other in the sky, making 
the background subtraction imprecise. Third, we discard any group with 
$M_{\rm h}<10^{13}M_{\odot}$ that has its center contained in the 
projection of a more massive group with $M_{\rm h}\ge 10^{13}M_{\odot}$. 
We use 200 bootstrap re-samplings to estimate measurement errors
of the conditional luminosity functions. 
The results of the CLF for galaxies in groups/halos of 
different masses are shown in \autoref{fig:clf}. Black data points represent 
results for satellite galaxies while grey data points for central galaxies. 

We use analytical models to fit the measured CLF, but for central and 
satellite galaxies separately. The total CLF at given halo mass can then be 
given by the sum of the CLF models of centrals and satellites.
For central galaxies, we use a Gaussian function,
\begin{equation}
    \Phi_c\left(M\right)=\frac{1}{\sqrt{2\pi}\sigma}
    \exp\left[\frac{-(M-\mu)^2}{2\sigma^2}\right]\,,
    \label{eq:gaussian}
\end{equation}
where $M$ is the absolute magnitude, and $\mu$ and $\sigma$ the mean
and dispersion of the distribution. We use a Monte Carlo Markov Chain (MCMC) 
method for the fitting, 
with each data point weighted by the uncertainty obtained from bootstrap
re-sampling. The fitting results are shown as the solid green lines 
in \autoref{fig:clf}, and the corresponding fitting parameters are listed 
in \autoref{tab:para_whole} for reference. The relation between 
$\mu$ and $M_{\rm h}$ obtained here is in general agreement with that 
obtained before \citep[e.g.][]{Yang2009CLF,Lan2016}.

To fit the CLF for satellite galaxies, 
we use different functional forms for the bright end ($M_{\rm r} < -18$) 
and the faint end ($M_{\rm r} \ge -18$). For the bright end, we use a
Schechter function, 
\begin{small}
\begin{equation}
    \Phi_{\rm b}\left(M\right)=N_{\rm b}\times 10^{-0.4\left(M-M^\ast_{\rm b}\right)\left(\alpha_{\rm b}+1\right)}
    \exp\left[-10^{-0.4\left(M-M^\ast_{\rm b}\right)}\right], 
    \label{eq:schechter}
\end{equation}
\end{small}
where $M^\ast_{\rm b}$ is a characteristic magnitude, $\alpha_{\rm b}$ 
the faint end slope and 
$N_{\rm b}$ an overall amplitude. For the faint end, we use a power law function,
\begin{equation}
    \Phi_{\rm f}(M)=N_{\rm f}\times 10^{-0.4\left(M-M^\ast_{\rm f}\right)
    	\left(\alpha_{\rm f}+1\right)}\,,
    \label{eq:powerlaw}
\end{equation}
where $M^\ast_{\rm f}$ is a characteristic magnitude, 
$\alpha_{\rm f}$ the power index and $N_{\rm f}$ an overall amplitude.
We find the fitting result is insensitive to the chosen value of 
$M^\ast_{\rm f}$, and so we fix it to be $M^\ast_{\rm f}=-18$ to reduce 
the number of free parameters. 
We use these functional forms to model the conditional luminosity function 
of satellite galaxies, instead of the double Schechter function adopted 
by \citetalias{Lan2016}, to reduce the degeneracy between model parameters.
We note that we obtained consistent results when using a double 
Schechter function to fit the data. The functional forms and the separation 
at $M_{\rm r}=-18$ are motivated by an inspection of the measured CLF, 
and the model fits are aimed at providing a compact description of the CLF 
measurements.

The fitting model adopted here contains five parameters: the bright end 
amplitude $N_{\rm b}$, the characteristic magnitude for the bright end 
$M^\ast_{\rm b}$, the slope for the bright end $\alpha_{\rm b}$, the faint end amplitude $N_{\rm f}$ 
and the faint end power index $\alpha_{\rm f}$. We require that 
$\Phi_{\rm b}$ and $\Phi_{\rm f}$ join continuously at $M_{\rm r} = -18$, 
so that we have a constraint on the model parameters:
\begin{equation}
    \Phi_{\rm b}(-18)=\Phi_{\rm f}(-18).
    \label{eq:constraint_indivi}
\end{equation}
Thus, the model is completely specified by four free parameters. Here again, we 
use the MCMC method to fit the measured CLF of satellite galaxies
to the model. The results are shown as green dashed lines in \autoref{fig:clf}
for groups/halos of different mass. As one can see, the functional forms adopted 
can fit the observational results very well. The best-fit values and the errors of 
the model parameters are shown in \autoref{fig:para_whole} and listed in 
\autoref{tab:para_whole}. The amplitudes of the 
CLF, $N_{\rm f}$ and $N_{\rm b}$, increase almost linearly with halo mass, 
as shown by the fitting results presented in the right panel.
The characteristic magnitude, $M_{\rm b}^\ast$, is almost independent of $M_{\rm h}$ at 
$M_{\rm h}>10^{13} {\rm M}_\odot$, but becomes significantly fainter
at lower $M_{\rm h}$. This $M_{\rm h}$-dependence of $M_{\rm b}^\ast$ is well 
described by the functional form indicated in the left panel.
Both of the slopes, $\alpha_{\rm b}$ and $\alpha_{\rm f}$, are quite independent of 
halo mass, and $\alpha_{\rm f}$ is significantly steeper than $\alpha_{\rm b}$
(see the middle panel). We notice that the faint-end slope, almost constant 
at $\alpha_{\rm f}=-1.59$, is very close to the faint-end slope of the luminosity 
function of the general population of low-$z$ galaxies which is $\alpha\sim-1.6$ 
after the cosmic variance in the SDSS sample is corrected 
\citep[][see their Figure 8]{Li2022}.

\begin{deluxetable*}{lccccccc}
\tabletypesize{\scriptsize}
\tablewidth{0pt}
\tablecaption{Model parameters of CLF for the whole sample \label{tab:para_whole}}
\tablehead{
\colhead{$\log\left(M_{\rm h}/M_{\odot}\right)$} & \colhead{$\mu$}& \colhead{$\sigma$} & \colhead{$N_{\rm b}$} & \colhead{$M_{\rm b}^{*}$} & \colhead{$\alpha_{\rm b}$} & \colhead{$N_{\rm f}$} & \colhead{$\alpha_{\rm f}$}}
\startdata
        \hline
		12.00$\sim$ 12.34 & $-21.199^{+0.002}_{-0.002}$ & $0.339^{+0.001}_{-0.001}$ &$0.47^{+0.03}_{-0.03}$ & $-19.70^{+0.23}_{-0.25}$ & $-0.61^{+0.26}_{-0.25}$ & $0.21^{+0.05}_{-0.09}$ & $-1.67^{+0.03}_{-0.03}$ \\
		12.34$\sim$ 12.68 & $-21.729^{+0.003}_{-0.003}$ & $0.314^{+0.003}_{-0.002}$ & $0.62^{+0.03}_{-0.03}$ & $-20.45^{+0.23}_{-0.27}$ & $-0.84^{+0.19}_{-0.17}$ & $0.40^{+0.11}_{-0.17}$ & $-1.66^{+0.04}_{-0.03}$ \\
		12.68$\sim$ 13.03 & $-22.150^{+0.005}_{-0.004}$ & $0.319^{+0.003}_{-0.003}$ & $0.76^{+0.03}_{-0.03}$ & $-21.46^{+0.40}_{-0.60}$ & $-1.03^{+0.16}_{-0.15}$ & $0.80^{+0.29}_{-0.58}$ & $-1.62^{+0.04}_{-0.04}$ \\
		13.03$\sim$ 13.37 & $-22.413^{+0.008}_{-0.007}$ & $0.368^{+0.005}_{-0.005}$ & $2.30^{+0.06}_{-0.06}$ & $-21.24^{+0.12}_{-0.13}$ & $-1.04^{+0.07}_{-0.07}$ & $2.47^{+0.42}_{-0.56}$ & $-1.57^{+0.03}_{-0.03}$ \\
		13.37$\sim$ 13.71 & $-22.601^{+0.011}_{-0.011}$ & $0.388^{+0.006}_{-0.007}$ & $2.84^{+0.08}_{-0.08}$ & $-21.76^{+0.19}_{-0.21}$ & $-1.14^{+0.08}_{-0.07}$ & $4.43^{+0.99}_{-1.33}$ & $-1.42^{+0.06}_{-0.05}$ \\
		13.71$\sim$ 14.05 & $-22.777^{+0.016}_{-0.015}$ & $0.429^{+0.021}_{-0.018}$ & $8.07^{+0.19}_{-0.20}$ & $-21.61^{+0.11}_{-0.11}$ & $-1.05^{+0.06}_{-0.06}$ & $9.18^{+1.52}_{-1.82}$ & $-1.57^{+0.03}_{-0.03}$ \\
		14.05$\sim$ 14.39 & $-22.946^{+0.025}_{-0.024}$ & $0.422^{+0.021}_{-0.020}$ & $19.46^{+0.53}_{-0.54}$ & $-21.53^{+0.09}_{-0.09}$ & $-0.96^{+0.06}_{-0.06}$ & $16.32^{+2.48}_{-2.85}$ & $-1.50^{+0.07}_{-0.06}$ \\
		14.39$\sim$ 14.73 & $-23.189^{+0.084}_{-0.115}$ & $0.610^{+0.137}_{-0.095}$ & $34.21^{+0.93}_{-0.93}$ & $-21.81^{+0.09}_{-0.10}$ & $-1.12^{+0.05}_{-0.05}$ & $50.73^{+7.52}_{-9.44}$ & $-1.40^{+0.06}_{-0.06}$ \\
		14.73$\sim$ 15.08 & $-23.710^{+0.152}_{-0.180}$ & $0.535^{+0.172}_{-0.106}$ & $92.37^{+3.47}_{-3.36}$ & $-21.66^{+0.12}_{-0.13}$ & $-1.03^{+0.08}_{-0.08}$ & $99.37^{+21.10}_{-27.68}$ & $-1.65^{+0.05}_{-0.04}$ \\
		\hline
\enddata
\end{deluxetable*}

\subsection{Dependence on galaxy color}
\label{sec:color}

\begin{figure*}
    \centering
	\includegraphics[width=0.8\textwidth]{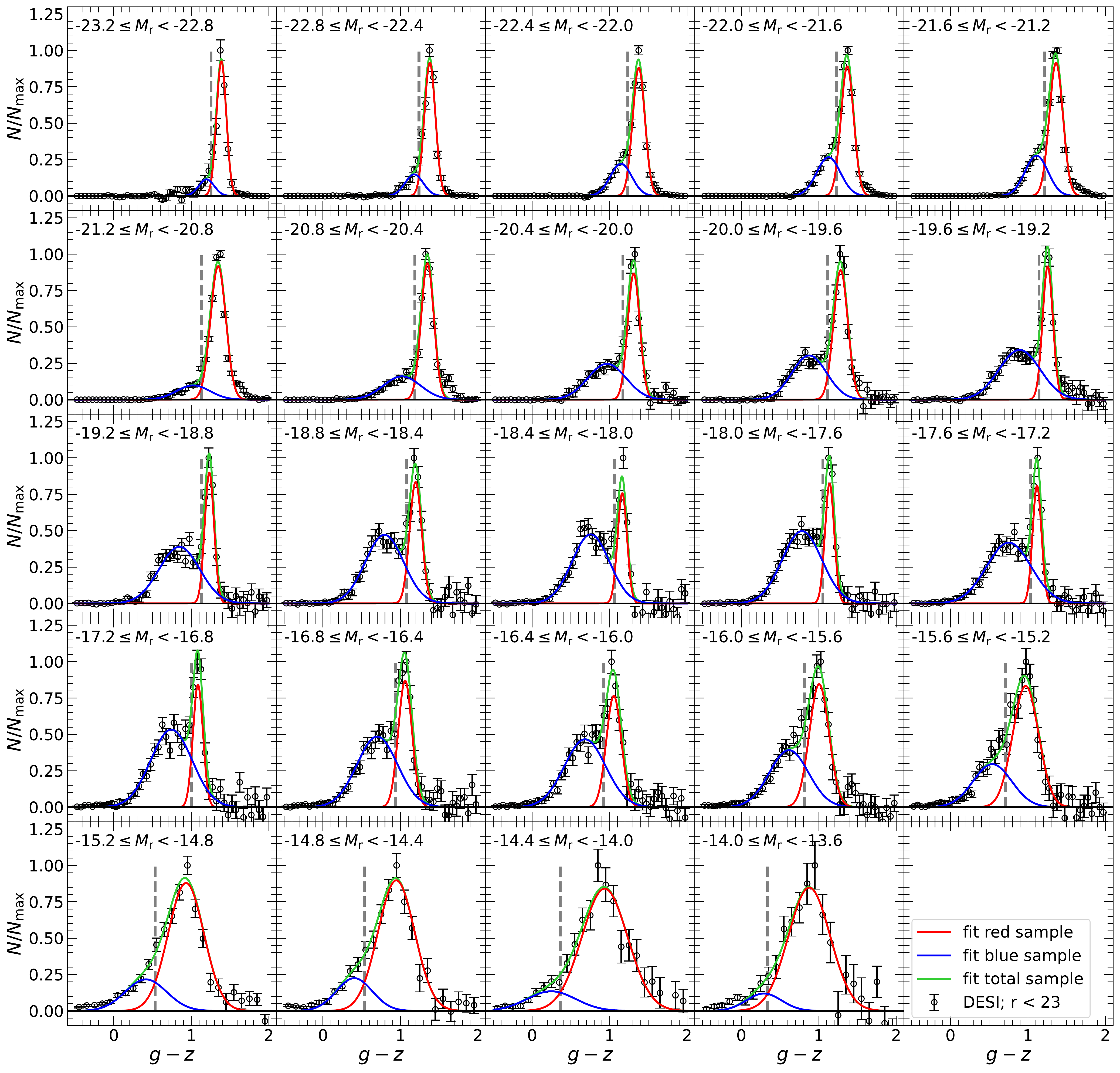}
    \caption{The black dots represent the $(g-z)$ color distributions of all member galaxies in the galaxy groups with $M_{\rm h}>10^{12}M_{\odot}$ at $0.01 < z < 0.08$ using DESI imaging data. The red and blue solid lines are the fitting results with Gaussian function for the red sequence and blue sequence. The green lines are the overall fitting results. The grey vertical dash lines are the dividing lines to separate the red and blue populations.}
    \label{fig:gz_dis}
\end{figure*}

\begin{figure}
    \centering
	\includegraphics[width=0.8\columnwidth]{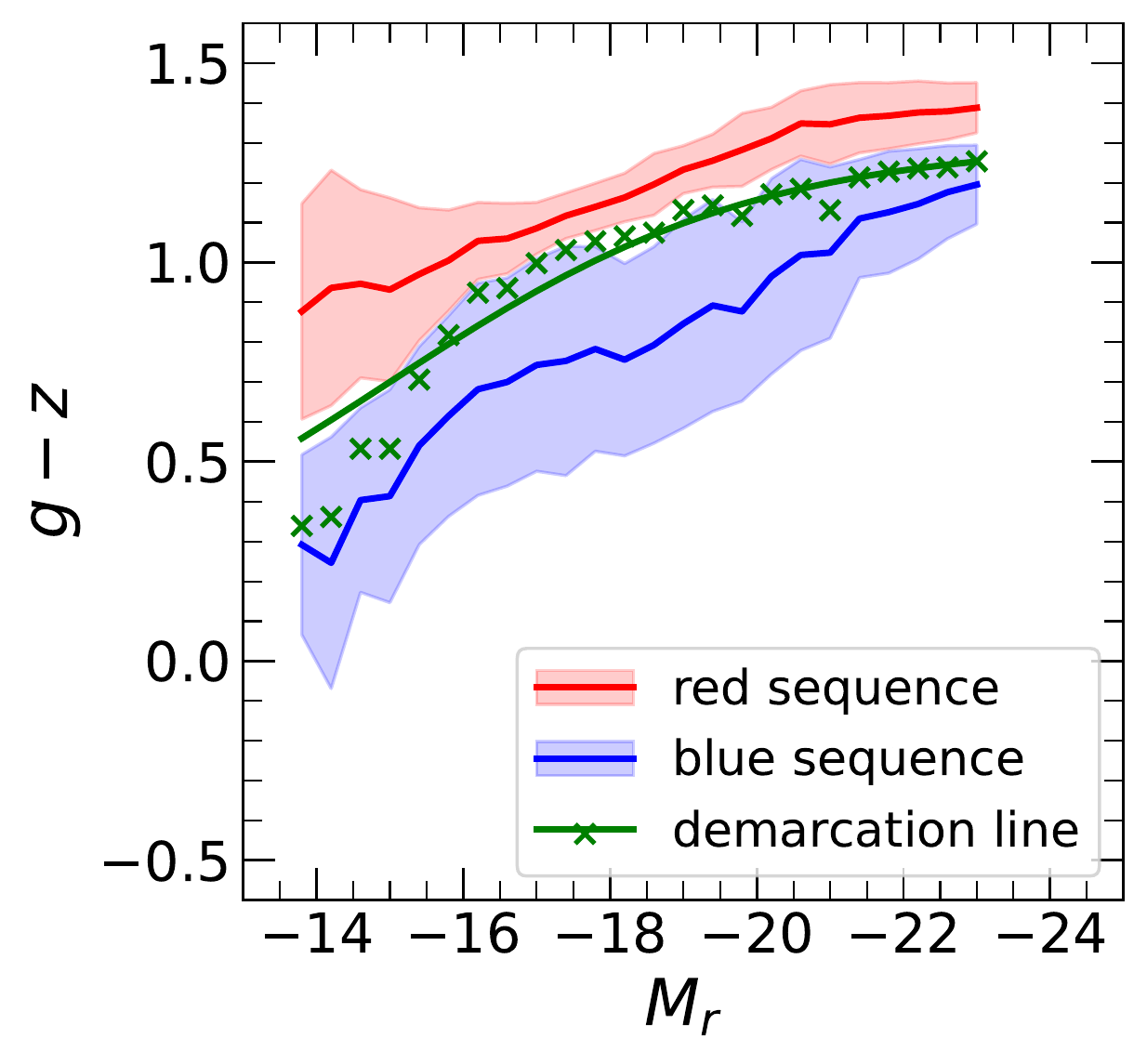}
    \caption{Here we show the fitted red and blue sequence in Fig.\,\ref{fig:gz_dis} for $(g-z)$ color as the function of absolute magnitude. The solid lines represent the mean value and the shade region represent the 1-$\sigma$ scatter of the sequence. The green crosses represent the demarcation points at given absolute magnitude. The green solid line are the fitted demarcation line of Eq.\,\ref{eq:demarcation}.}
    \label{fig:gz_sequence}
\end{figure}

\begin{figure*}
    \centering
    \includegraphics[scale=0.29]{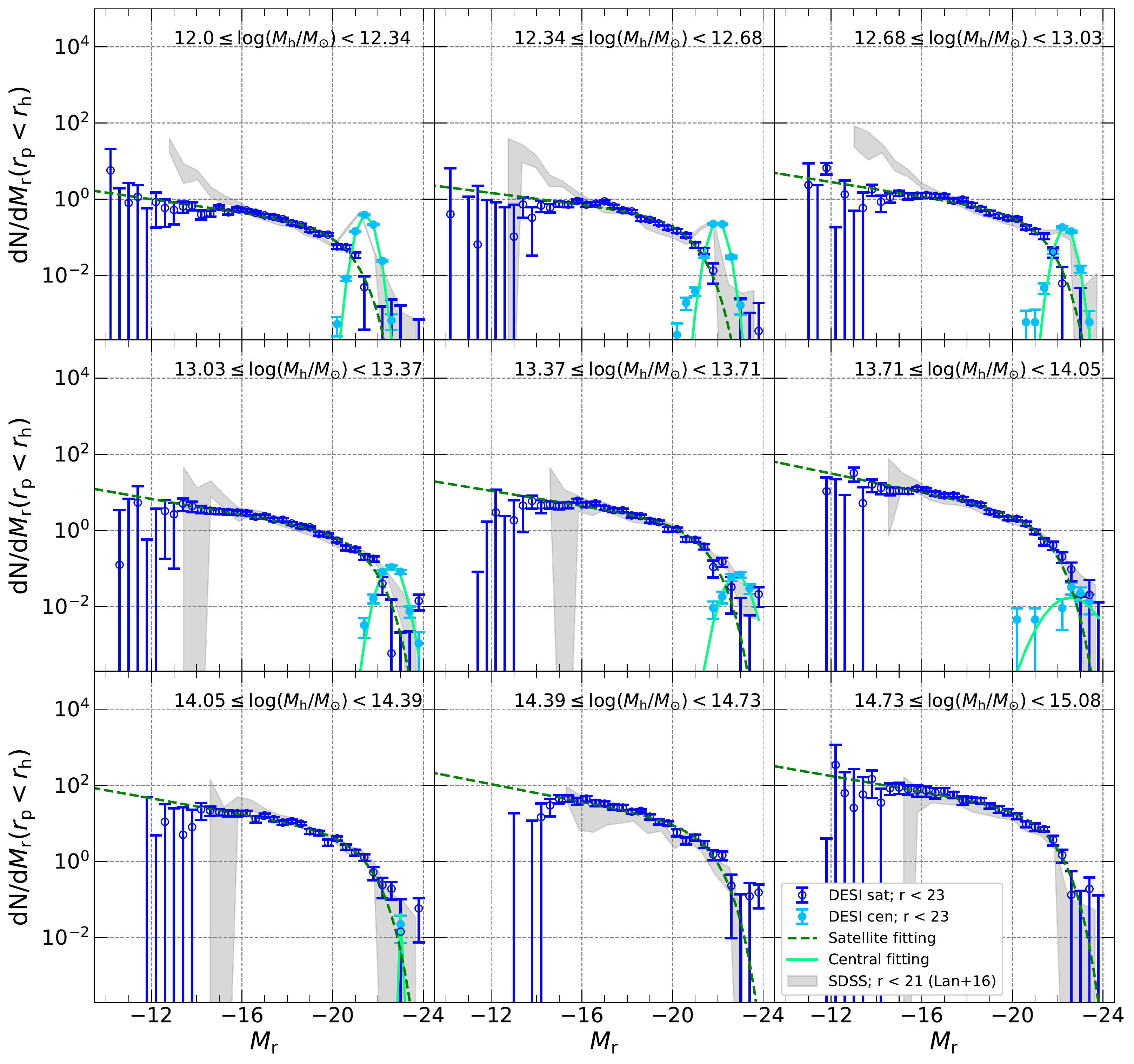}
    \caption{The dark blue hollow dots and light blue solid dots represent our measurements of the luminosity functions for central 
    galaxies and satellite galaxies of blue sample in galaxy groups at $0.01 \le z \le 0.08$ with the DESI imaging data. 
    The grey shaded regions represent the results of total luminosity functions in galaxy groups for the blue sample 
    in the redshift range $0.01 < z < 0.05$ obtained by \citetalias{Lan2016} using SDSS imaging data. 
    The green solid and dash lines are the 
    fitting results for our central and satellite luminosity functions respectively.}
    \label{fig:clf_blue}
\end{figure*}

\begin{figure*}
    \centering
	\includegraphics[scale=0.29]{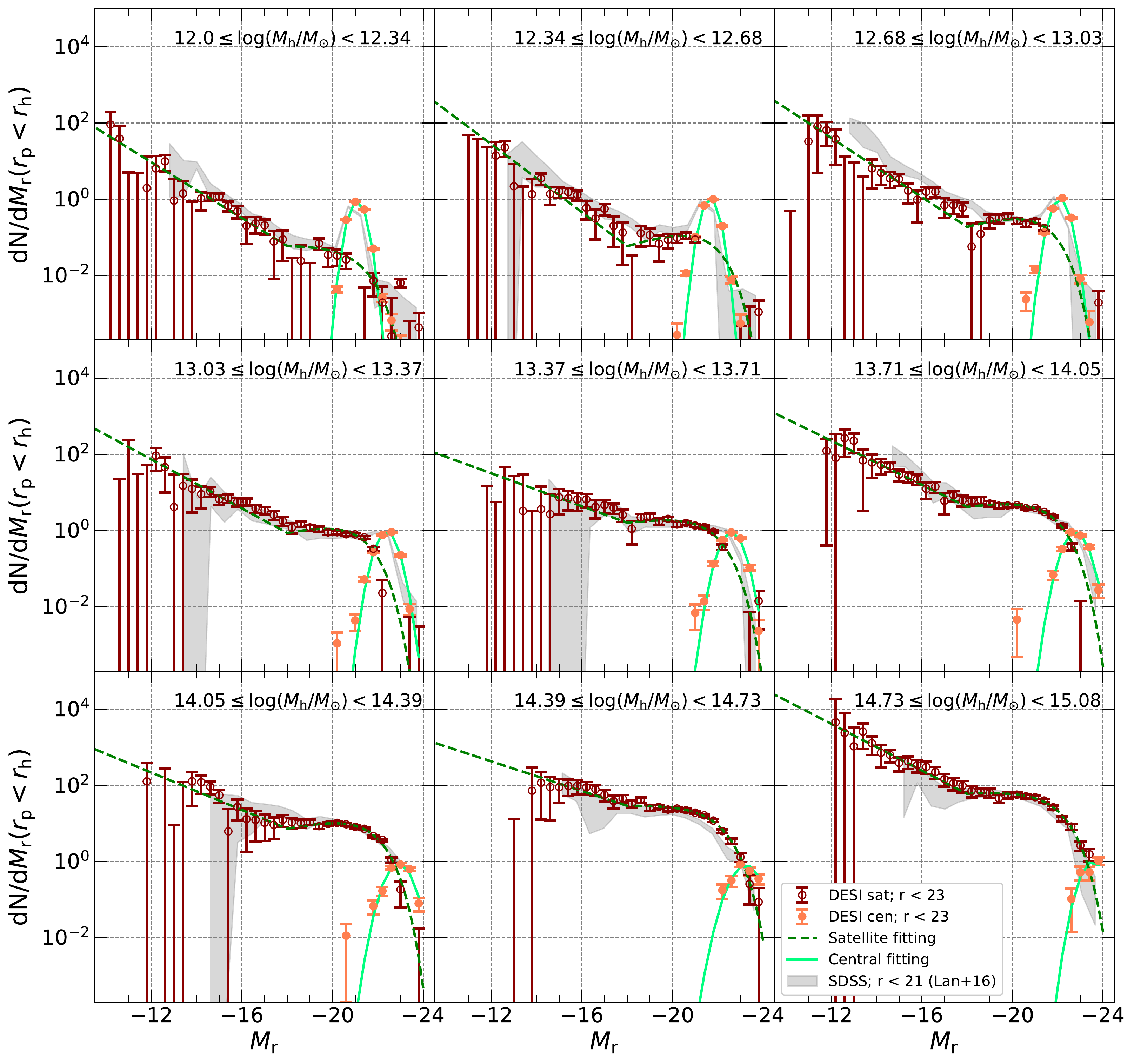}
    \caption{The dark red hollow dots and light red solid dots represent our measurements of the luminosity functions for central 
    galaxies and satellite galaxies of red sample in galaxy groups at $0.01 \le z \le 0.08$ with the DESI imaging data. 
    The grey shaded regions represent the results of total luminosity functions in galaxy groups for the red sample 
    in the redshift range $0.01 < z < 0.05$ obtained by \citetalias{Lan2016} using SDSS imaging data. 
    The green solid and dash lines are the 
    fitting results for our central and satellite luminosity functions respectively.}
    \label{fig:clf_red}
\end{figure*}

\begin{figure*}
    \centering
    \includegraphics[width=0.75\textwidth]{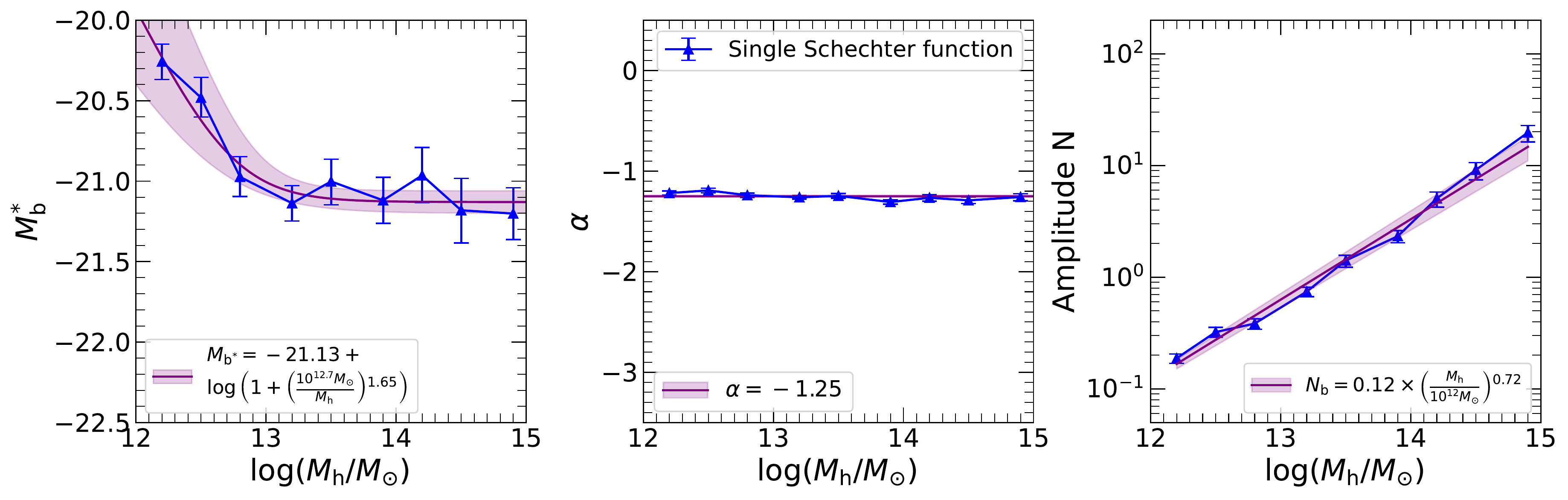}
	\includegraphics[width=0.75\textwidth]{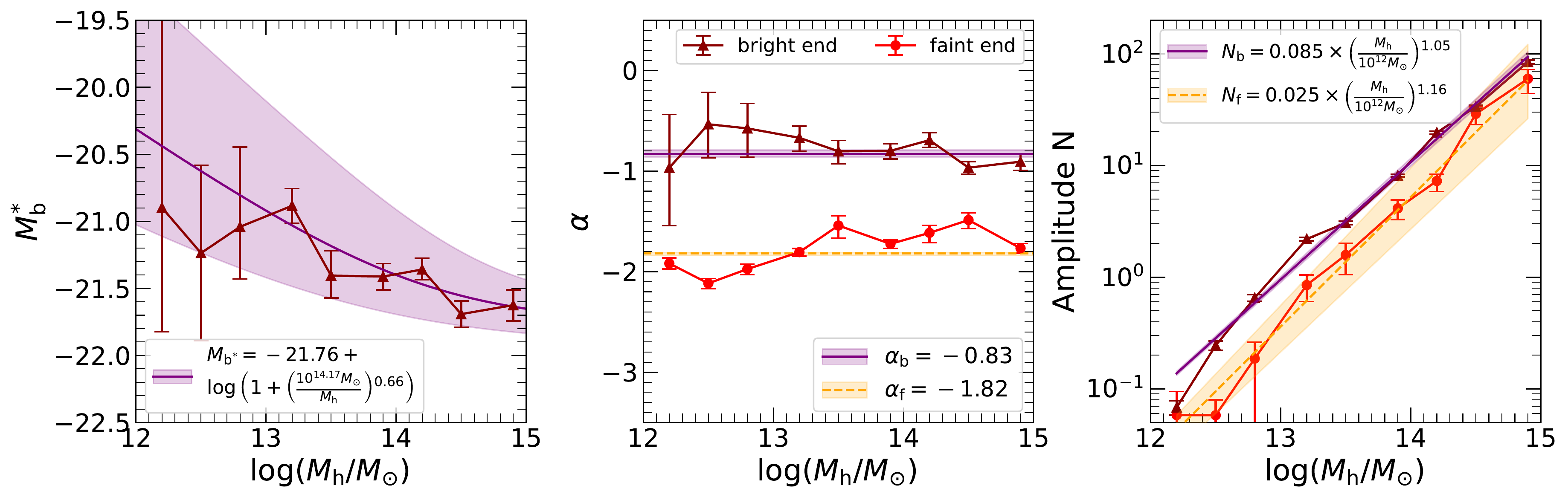}
    \caption{The figures show the model parameters for fitting the conditional luminosity functions of the red sample and blue sample as the function of halo mass. The errorbars represent the 1-$\sigma$ scatters of the parameters. We choose the same function forms as in Fig.\,\ref{fig:para_whole} to describe the dependence of the parameters on the halo mass. The functions are written in the figures, shown as the purple and orange lines for the bright and faint end respectively.}
    \label{fig:para_color}
\end{figure*}

In this subsection, we examine the rest-frame color distribution 
as a function of absolute magnitude for member galaxies of groups/halos. 
In this case, the galaxy property vector is 
$\vec{\mathbf{q}}=\{M_{\rm r}, g-z\}$, where the $(g-z)$ color can be obtained 
from DESI-DR9. Let us first look at all member galaxies in groups 
with $M_{\rm h}\ge 10^{12}M_{\odot}$ at $0.01\le z\le0.08$. The black data points 
in \autoref{fig:gz_dis} show the $(g-z)$ color distributions for group galaxies 
in different intervals of $M_{\rm r}$. As one can see, over a large range of 
$M_{\rm r}$, the color distributions are bimodal. We use a bi-Gaussian 
function to fit each of the distributions, and 
the fitting results are shown as colored curves in \autoref{fig:gz_dis}. 
The red and blue curves are for red and blue modes, respectively, 
while the green curves are the sum of the two. It is clear that a
bi-Gaussian function can fit the observational data very well, and we can use the 
fitting results to divide galaxies into two populations. 
For example, one may use the point where the two Gaussian curves intersect 
to separate galaxies into red and blue populations for each $M_{\rm r}$ 
bin. This separation is marked by the vertical dashed line in each panel of 
\autoref{fig:gz_dis}. Qualitatively, as the luminosity of galaxies decreases, 
the red fraction first decreases, reaching the lowest value at $M_{\rm r}\sim -18$, 
and then increases. In the faint end ($M_{\rm r}>-15$), almost all galaxies 
are contained in the red population. We will discuss the results in more detail in 
\autoref{sec:old_population}. 

In \autoref{fig:gz_sequence}, we plot the $(g-r)$ color as a function of $M_{\rm r}$. 
The red and blue lines represent the mean values of the Gaussian fitting for the two modes, 
while the shaded regions represent the corresponding 1-$\sigma$ ranges. 
The green crosses are the separation points shown by the vertical lines in 
\autoref{fig:gz_dis}. We also obtain a demarcation line following \citet{Baldry2004}:
\begin{equation}
    T(M_{\rm r})=p_0+q_0\tanh\left(\frac{M_{\rm r}-q_1}{q_2}\right).
\end{equation}
To reduce degeneracy between parameters, we set $q_1=-15$ and $q_2=5$, 
treating only $p_0$ and $q_0$ as free parameters. For a given 
choice of demarcation (i.e. for given $p_0$ and $q_0$), we calculate 
a `completeness' factor, ${\cal C}_{\rm r}$ or ${\cal C}_{\rm b}$,  
which is defined as the fraction of galaxies in the red or blue mode that 
are selected according to the demarcation line, 
and a `reliability' factor, ${\cal R}_{\rm r}$ or ${\cal R}_{\rm b}$, 
defined as the fraction among all galaxies separated into the red or blue 
mode by the demarcation line that are truly red or blue mode galaxies. 
These two factors are estimated for each of the $M_{\rm r}$-samples, 
$i$, and we compute the value of 
\begin{equation}
{\cal P}\equiv 
\prod_{\rm i}\mathcal{C}_{\rm r,i}\mathcal{R}_{\rm r,i}\mathcal{C}_{\rm b,i}\mathcal{R}_{\rm b,i}
\end{equation}
for the set of parameters in question, and maximize ${\cal P}$ by 
varying the values of $p_0$ and $q_0$. 
The optimized color demarcation line so obtained is 
\begin{equation}
    T(M_{\rm r})=0.7-0.6\tanh\left(\frac{M_{\rm r}+15}{5}\right),
    \label{eq:demarcation}
\end{equation}
which is shown as the green line in \autoref{fig:gz_sequence}. 
This demarcation line matches that represented by the green 
crosses well at $M_{\rm r} < -15$. From \autoref{fig:gz_dis} we can see that 
both the red and blue modes become broad and overlap with each other
at $M_{\rm r} > -15$, making it difficult to have a clean separation of the two 
modes. However, our tests showed that this uncertainty in the color demarcation 
does not have a significant impact on our results.

We use \autoref{eq:demarcation} to divide group galaxies 
into red and blue populations and compute the CLF 
separately for them. The results are shown in \autoref{fig:clf_blue}
and \autoref{fig:clf_red} for the blue and red populations, respectively.  
The light blue (red) and dark blue (red) data points represent the results 
of central and satellite galaxies in the blue (red) population. For blue satellites, 
the CLF can be well fitted by a single Schechter function. The fitting results are 
shown as dashed green lines in \autoref{fig:clf_blue}, and the corresponding 
fitting parameters are shown as functions of halo mass in the upper panels of 
\autoref{fig:para_color} and are listed in \autoref{tab:para_color} for reference. 
As one can see, the single Schechter model describes the data well. 

We use the same power-law plus Schechter model described in \autoref{sec:clf}
to fit the CLF for red satellites, and the fitting results are shown 
as dashed green lines in \autoref{fig:clf_red}. 
The fitting parameters are plotted as functions of halo mass in the lower panels of 
\autoref{fig:para_color} and are listed in \autoref{tab:para_color} for reference. 
We again see that the fitting model describes well all the observed CLF of the red 
population. For reference, we use the same functional forms described in
\autoref{sec:clf} to fit the dependence of the model parameters on halo mass. 
The results are shown in \autoref{fig:para_color} and the corresponding fitting 
functions are provided in each panel. 

As one can see from \autoref{fig:para_color}, the faint-end slope for the CLF
of the blue population, $\alpha\approx -1.25$, is almost completely independent of 
halo mass. This is also true for the brighter (Schechter) portion of the CLF of the 
red population, although the slope, $\alpha_{\rm b}\approx 0.8$ is much shallower. 
The faint, power-law portion of the CLF for the red population has a slope, 
$\alpha_{\rm f}\sim -1.8$, although there are some fluctuations from sample to 
sample. For the blue population, the characteristic 
absolute magnitude is roughly a constant, $M^\ast_{\rm{b}}\sim -21.1$, for halos with 
$M_{\rm h}> 10^{13}M_\odot$, and increases (becomes fainter) with decreasing 
$M_{\rm h}$ at $M_{\rm h}< 10^{13}M_\odot$. The dependence of $M^\ast_{\rm{b}}$ 
on halo mass for the Schechter portion of the red population is roughly the same 
as the blue population, although the errors in the measurements are much larger at 
the low-$M_{\rm h}$ end. At a given $M_{\rm h}$, the characteristic absolute magnitude 
for the red population is in general brighter than that for the blue population. Finally, 
the overall amplitude of the CLF is roughly a power-law for both populations, 
but it is sub-linear, with a power index of $\sim 0.7$, for blue galaxies, 
and slightly super-linear, with a power index of $\sim 1.1$, for the red population.  
These results are consistent with those obtained previously, as we will discuss 
in the following subsection.

\begin{deluxetable*}{lccccccc}
\tabletypesize{\scriptsize}
\tablewidth{0pt}
\tablecaption{Model parameters of CLF for red and blue sample \label{tab:para_color}}
\tablehead{
\colhead{$\log\left(M_{\rm h}/M_{\odot}\right)$} & \colhead{$\mu$}& \colhead{$\sigma$} & \colhead{$N_{\rm b}$} & \colhead{$M_{\rm b}^{*}$} & \colhead{$\alpha_{\rm b}$} & \colhead{$N_{\rm f}$} & \colhead{$\alpha_{\rm f}$}}
\startdata
        \hline
		red \\
		12.00$\sim$ 12.34 & $-21.085^{+0.002}_{-0.002}$ & $0.296^{+0.001}_{-0.001}$ & $0.06^{+0.01}_{-0.01}$ & $-20.90^{+0.92}_{-1.60}$ & $-0.97^{+0.58}_{-0.53}$ & $0.06^{+0.04}_{-0.39}$ & $-1.92^{+0.06}_{-0.06}$ \\
		12.34$\sim$ 12.68 & $-21.666^{+0.003}_{-0.003}$ & $0.298^{+0.003}_{-0.003}$ & $0.25^{+0.02}_{-0.02}$ & $-21.24^{+0.65}_{-0.66}$ & $-0.53^{+0.33}_{-0.32}$ & $0.06^{+0.02}_{-0.07}$ & $-2.12^{+0.05}_{-0.05}$ \\
		12.68$\sim$ 13.03 & $-22.113^{+0.005}_{-0.005}$ & $0.309^{+0.003}_{-0.003}$ & $0.65^{+0.04}_{-0.04}$ & $-21.04^{+0.39}_{-0.60}$ & $-0.58^{+0.29}_{-0.25}$ & $0.19^{+0.07}_{-0.14}$ & $-1.97^{+0.06}_{-0.06}$ \\
		13.03$\sim$ 13.37 & $-22.392^{+0.007}_{-0.008}$ & $0.356^{+0.005}_{-0.005}$ & $2.20^{+0.08}_{-0.08}$ & $-20.88^{+0.13}_{-0.13}$ & $-0.67^{+0.13}_{-0.12}$ & $0.85^{+0.20}_{-0.25}$ & $-1.81^{+0.04}_{-0.04}$ \\
		13.37$\sim$ 13.71 & $-22.592^{+0.011}_{-0.011}$ & $0.391^{+0.007}_{-0.007}$ & $3.06^{+0.11}_{-0.11}$ & $-21.40^{+0.17}_{-0.18}$ & $-0.80^{+0.12}_{-0.11}$ & $1.58^{+0.42}_{-0.53}$ & $-1.54^{+0.12}_{-0.10}$ \\
		13.71$\sim$ 14.05 & $-22.778^{+0.017}_{-0.016}$ & $0.413^{+0.012}_{-0.012}$ & $8.13^{+0.21}_{-0.23}$ & $-21.41^{+0.10}_{-0.10}$ & $-0.80^{+0.08}_{-0.07}$ & $4.14^{+0.78}_{-0.86}$ & $-1.72^{+0.04}_{-0.04}$ \\
		14.05$\sim$ 14.39 & $-22.943^{+0.028}_{-0.026}$ & $0.429^{+0.022}_{-0.021}$ & $19.68^{+0.57}_{-0.57}$ & $-21.36^{+0.08}_{-0.08}$ & $-0.69^{+0.07}_{-0.07}$ & $7.26^{+1.12}_{-1.43}$ & $-1.61^{+0.10}_{-0.08}$ \\
		14.39$\sim$ 14.73 & $-23.182^{+0.079}_{-0.095}$ & $0.594^{+0.140}_{-0.093}$ & $33.64^{+0.99}_{-1.00}$ & $-21.69^{+0.10}_{-0.10}$ & $-0.97^{+0.06}_{-0.06}$ & $29.01^{+4.84}_{-5.84}$ & $-1.49^{+0.09}_{-0.07}$ \\
		14.73$\sim$ 15.08 & $-23.720^{+0.170}_{-0.201}$ & $0.553^{+0.202}_{-0.117}$ & $84.59^{+3.24}_{-3.20}$ & $-21.62^{+0.12}_{-0.11}$ & $-0.91^{+0.09}_{-0.08}$ & $59.75^{+12.42}_{-15.58}$ & $-1.77^{+0.05}_{-0.05}$ \\
		\hline
        blue \\
		12.00$\sim$ 12.34 & $-21.455^{+0.004}_{-0.004}$ & $0.314^{+0.003}_{-0.003}$ & $0.19^{+0.02}_{-0.02}$ & $-20.26^{+0.11}_{-0.11}$ & $-1.22^{+0.02}_{-0.02}$ \\
		12.34$\sim$ 12.68 & $-21.995^{+0.006}_{-0.006}$ & $0.289^{+0.005}_{-0.005}$ & $0.32^{+0.03}_{-0.03}$ & $-20.48^{+0.12}_{-0.13}$ & $-1.19^{+0.02}_{-0.02}$ \\
		12.68$\sim$ 13.03 & $-22.330^{+0.012}_{-0.012}$ & $0.300^{+0.009}_{-0.009}$ & $0.38^{+0.04}_{-0.04}$ & $-20.97^{+0.12}_{-0.13}$ & $-1.24^{+0.02}_{-0.02}$ \\
		13.03$\sim$ 13.37 & $-22.566^{+0.022}_{-0.023}$ & $0.388^{+0.015}_{-0.016}$ & $0.74^{+0.08}_{-0.07}$ & $-21.14^{+0.11}_{-0.11}$ & $-1.26^{+0.02}_{-0.02}$ \\
		13.37$\sim$ 13.71 & $-22.865^{+0.057}_{-0.063}$ & $0.453^{+0.081}_{-0.058}$ & $1.41^{+0.18}_{-0.16}$ & $-21.00^{+0.15}_{-0.14}$ & $-1.25^{+0.02}_{-0.02}$ \\
		13.71$\sim$ 14.05 & $-22.561^{+0.184}_{-0.187}$ & $0.783^{0.182}_{-0.377}$ & $2.33^{+0.30}_{-0.28}$ & $-21.13^{+0.14}_{-0.14}$ & $-1.31^{+0.02}_{-0.02}$ \\
		14.05$\sim$ 14.39 & $-23.039^{+0.077}_{-0.077}$ & $0.077^{+0.039}_{-0.038}$ & $5.08^{+0.85}_{-0.70}$ & $-20.96^{+0.17}_{-0.17}$ & $-1.27^{+0.03}_{-0.03}$ \\
		14.39$\sim$ 14.73 & & &$9.14^{+1.70}_{-1.46}$ & $-21.18^{+0.21}_{-0.20}$ & $-1.29^{+0.03}_{-0.03}$ \\
		14.73$\sim$ 15.08 & & & $19.74^{+3.47}_{-3.08}$ & $-21.20^{+0.16}_{-0.16}$ & $-1.26^{+0.04}_{-0.03}$ \\
		\hline
\enddata
\end{deluxetable*}

\subsection{Comparison with previous results}
\label{sec:comparison}

\begin{figure*}
    \centering
	\includegraphics[width=0.75\textwidth]{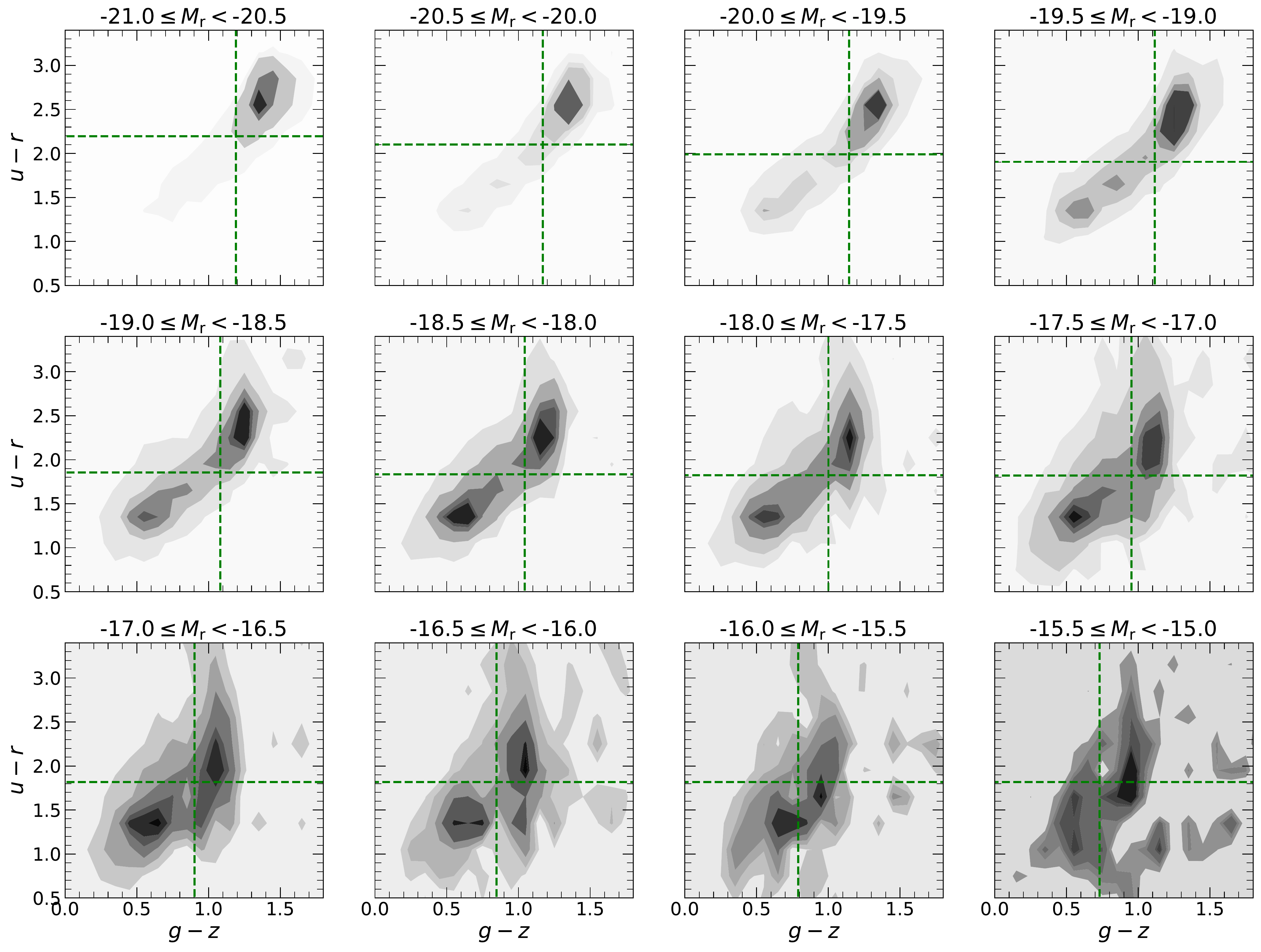}
    \caption{The color-color diagram in $(g-z)$-$(u-r)$ space for all member galaxies in the groups with DESI imaging data at different absolute magnitude. The green dash lines are the demarcation lines to separate the red and blue mode. The demarcation lines for $(g-z)$ color are from \autoref{eq:demarcation}. The demarcation lines for $(u-r)$ color are from \citet{Baldry2004}. }
    \label{fig:gz-ur_dis}
\end{figure*}

For comparison, the shaded grey regions in \autoref{fig:clf}, \autoref{fig:clf_blue} 
and \autoref{fig:clf_red} show the CLF measurements obtained by \citetalias{Lan2016} 
from the SDSS imaging data, and the shaded orange regions in \autoref{fig:clf}
show the measurements by \citet{Tinker2021} from the DESI DR6 and DR7 
imaging data. As one can see, our measurements match both of the previous 
results. For the total CLFs, the only exception is seen at 
$14.39\leq \log(M_{\rm h}/M_\odot)<14.73$, where the CLF obtained by \citet{Tinker2021} 
is higher at $M_{\rm r}>-19$. 
The reason is that \citet{Tinker2021} included all massive groups 
in this mass bin, but \citetalias{Lan2016} and 
our work have one more mass bin for the most massive groups as shown 
in the bottom-right panel. Since DESI-DR9 is about 2 magnitudes deeper than 
the SDSS data and DESI-DR9 has more pass times than the DESI-DR6/7, 
our measurements have smaller uncertainties than those obtained 
by \citetalias{Lan2016} and \citet{Tinker2021}. 

Another discrepancy can be seen in the low-luminosity ends ($M_{\rm r}>-16$) 
of the CLF in low-mass halos ($M_{\rm h}<10^{13}M_\odot$),
particularly for the blue population, where a significant upturn is seen in 
\citetalias{Lan2016} but absent in ours. 
This might be attributed to the fact that 
\citetalias{Lan2016} used the SDSS $(u-r)$ color, instead of the $(g-z)$ color used here, 
to separate galaxies into blue and red populations. In \autoref{fig:gz-ur_dis}, 
we show the distribution of group galaxies in the $(u-r)$ versus $(g-z)$ diagram, 
where the $(u-r)$ color is from the SDSS photometry used in \citetalias{Lan2016} while 
the $(g-z)$ color is from the DESI-DR9 photometry used in our analyses. 
Group members are identified from the DESI-DR9 using the method described in 
\autoref{sec:method} and are matched to SDSS galaxies using their 
centroid positions (RA, Dec). Only galaxies 
brighter than the SDSS magnitude limit of $r<21$ are used. For comparison, 
the horizontal lines show the demarcations of red and blue populations 
adopted by \citetalias{Lan2016}, while the vertical lines are those adopted in 
the present paper. For $M_{\rm r}<-17$, the distribution is bimodal in both $(u-r)$ and 
$(g-z)$, and the demarcations adopted by \citetalias{Lan2016} and us give similar 
separations of the two populations. For fainter galaxies with $M_{\rm r}>-17$, 
a bimodal distribution can still be seen in the $(g-z)$ color and is 
delineated by our demarcation lines, but it is much less clear in the 
$(u-r)$ distribution. The demarcations adopted by \citetalias{Lan2016}
cut across the red modes, which might explain the strong upturn in 
the faint end of the CLF obtained by \citetalias{Lan2016} for blue galaxies in 
low-mass groups. There are also some differences in the fitting parameters 
between our results and those of \citetalias{Lan2016}. For low-mass halos, 
the characteristic magnitudes of the CLF from our fitting are slightly fainter. 
The bright ends of the CLF for satellite galaxies in these low-mas halos 
are quite noisy. The fitting results of \citetalias{Lan2016} sometimes extend to 
the range of $M_{\rm r}$ where the measurements of the CLF 
for satellite galaxies drop rapidly to zero, while our fitting results 
follow this rapid drop more closely. We note that there is degeneracy 
among model parameters, and that our fitting functions can  
match the measurements from the SDSS imaging data, as one can see in 
\autoref{fig:clf}, \autoref{fig:clf_blue} and \autoref{fig:clf_red}.

\section{The old and young populations of satellite galaxies}
\label{sec:old_population}

\begin{figure*}
    \centering
	\includegraphics[width=0.75\textwidth]{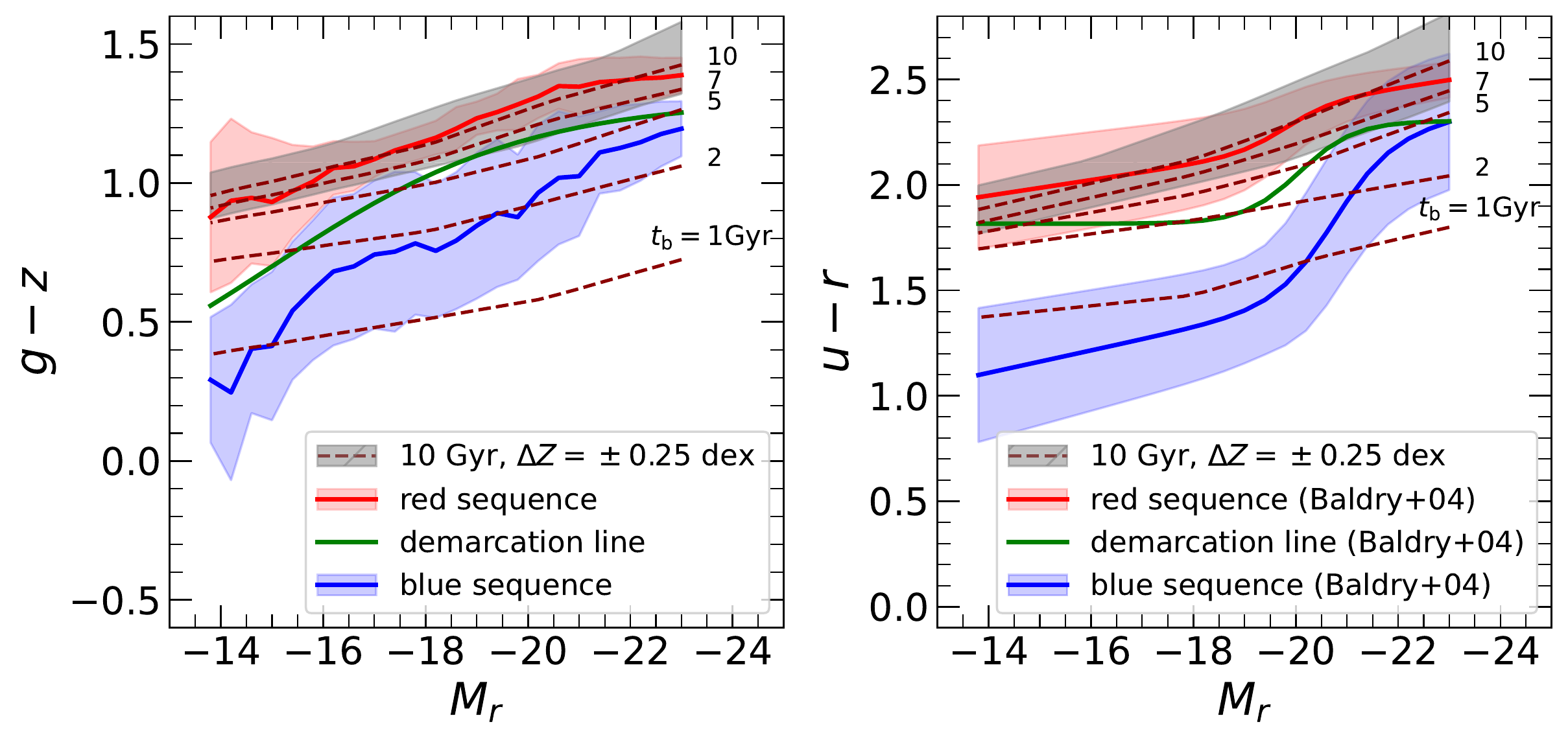}
    \caption{The figures show the red and blue sequence as the function of absolute magnitude for $(g-z)$ and $(u-r)$ color. The sequences for $(g-z)$ are same as \autoref{fig:gz_sequence} and the sequences for $(u-r)$ are from \citet{Baldry2004}. The dark red lines represent the results of the single burst model. The look-back time of the star burst are labelled on the figure.}
    \label{fig:color_sequence}
\end{figure*}

\subsection{The presence of an old sequence}
\label{sec:old_sequence}

As shown in \autoref{fig:gz_dis}, group galaxies have bimodal 
distributions in their $(g-z)$ colors.
At the very bright end, the population is dominated by galaxies in 
a narrow $(g-z)$ color distribution in the red component.
As galaxy luminosity decreases, the peak position of the red
component of the distribution shifts gradually towards the blue 
end, with the width remaining roughly constant for $M_{\rm r}<-16$
and becoming broader for fainter galaxies. Starting from the 
brightest end, the blue component becomes more important for fainter 
galaxies. The importance of the blue component reaches its maximum 
at $M_{\rm r}\sim -18$ and decreases gradually towards the fainter end. 
We note that the red and blue modes start to merge at $M_{\rm r}>-16$,
and the blue modes are observed only as an extended tail at the blue 
side of the $(g-z)$ distribution. We also note that the red modes 
for faint galaxies are actually bluer than the blue modes for galaxies 
at the brightest end. Thus, our separation of red and blue populations 
is not in terms of absolute color, but based on the bimodal distribution 
of galaxies of similar luminosity.  

To gain insights into potential causes of the color bimodality of group galaxies,  
we use the BC03 spectral synthesis model \citep{BC03} assuming the Chabrier 
initial mass function \citep{Chabrier2003} to provide some guidance. 
We first consider a simple case where the star formation history is a single 
burst at some look-back time $t_{\rm b}$:
\begin{equation}
    \Psi\left(t\right)\propto \delta\left(t-t_{\rm b}\right),
\end{equation}
where $t$ is the cosmic look-back time. The spectrum of a stellar population 
depends not only on the stellar age, but also on the stellar metallicity. 
We use the observed mass-metallicity relation (MZR) given in \citet{Kirby2013}
to model the metallicity: 
\begin{small}
\begin{equation}
    \log \left(\frac{Z}{Z_{\odot}}\right)=\left(-1.69\pm 0.04\right)+\left(0.30\pm 0.02\right)
    \log\left(\frac{M_\ast}{10^{6}M_{\odot}}\right),
    \label{eq:MZR}
\end{equation}
\end{small}
where $Z_{\odot}=0.02$ and $M_{\odot}$ are the solar metallicity and mass, respectively.
Note that the errors quoted here are uncertainties in the mean relation. The scatter in 
$\log (Z)$ at given $M_\ast$ is about $0.25\, {\rm dex}$.
To obtain the stellar mass of a galaxy from its luminosity, we use the relation between 
the mass-to-light ratio and color given in \citet{Bell2003}: 
\begin{small}
\begin{equation}
    \log\left(\frac{M_\ast}{M_\odot}\right)=-0.22 + 0.51\times(g-z) - 0.40 
    \times(M_{\rm r} - 4.64) - 0.15\,,
\end{equation}
\end{small}
and 
\begin{small}
\begin{equation}
    \log\left(\frac{M_\ast}{M_\odot}\right)=-0.22 + 0.30\times(u-r) - 0.40 \times(M_{\rm r} - 4.64) - 0.15.
\end{equation}
\end{small}
The single burst model is expected to be valid only for galaxies dominated by 
old stellar populations. We thus use the $(g-z)$ and $(u-r)$ 
colors of the red sequence shown in \autoref{fig:color_sequence}
to convert luminosity to stellar mass. We then use \autoref{eq:MZR}
to obtain the stellar metallicity. The BC03 model only provides 
spectra for stellar populations with metallicity $Z=0.0001, 0.0004, 0.004, 0.008, 0.02$ and 
$0.05$. To predict the color of the stellar population at the metallicity obtained from 
the mass-metallicity relation, we first obtain the $(g-z)$ and $(u-r)$ colors of 
the stellar populations of different ages for each of the 6 metallicities, we
then use a linear interpolation to obtain the two colors corresponding to  
the metallicity in question. 

The color-magnitude relations predicted by the single burst model with different 
ages are presented in \autoref{fig:color_sequence} as the dashed lines, 
with the left and right panels showing results for the $(g-z)$ and $(u-r)$ colors, 
respectively. The dark shade around the $10\,{\rm Gyr}$ line in each panel  
indicates the range covered by the variance of $\pm0.25{\rm dex}$ in the metallicity.  
Also included in the figure are observed color-magnitude relations for both the red and blue sequences 
obtained in \autoref{sec:color} for the $(g-z)$ versus $M_{\rm r}$ relation, and by 
\citet{Baldry2004} for the $(u-r)$ versus $M_{\rm r}$ relation. 
The green lines are the demarcation lines between the red and blue modes. 
As one can see, the red sequences in both $(g-z)$ and $(u-r)$ are consistent 
with the prediction of a single burst with an old age $\sim 10\,{\rm Gyr}$. 
Both the slope and width of the red sequence follow the prediction 
of the observed mass-metallicity relation, indicating that the red sequence is dominated 
by galaxies with an old stellar population, quite independent of the stellar mass of galaxies.   
The demarcation lines divide the galaxy population roughly at an age around 
$5\,{\rm Gyr}$. However, the shapes of the demarcation lines are quite 
different from those of the color-magnitude relations predicted by the 
single burst model, indicating that the star formation history of galaxies
becomes more complicated and likely mass-dependent as one moves to the blue 
sequence. 

\begin{figure*}
    \centering
	\includegraphics[width=0.75\textwidth]{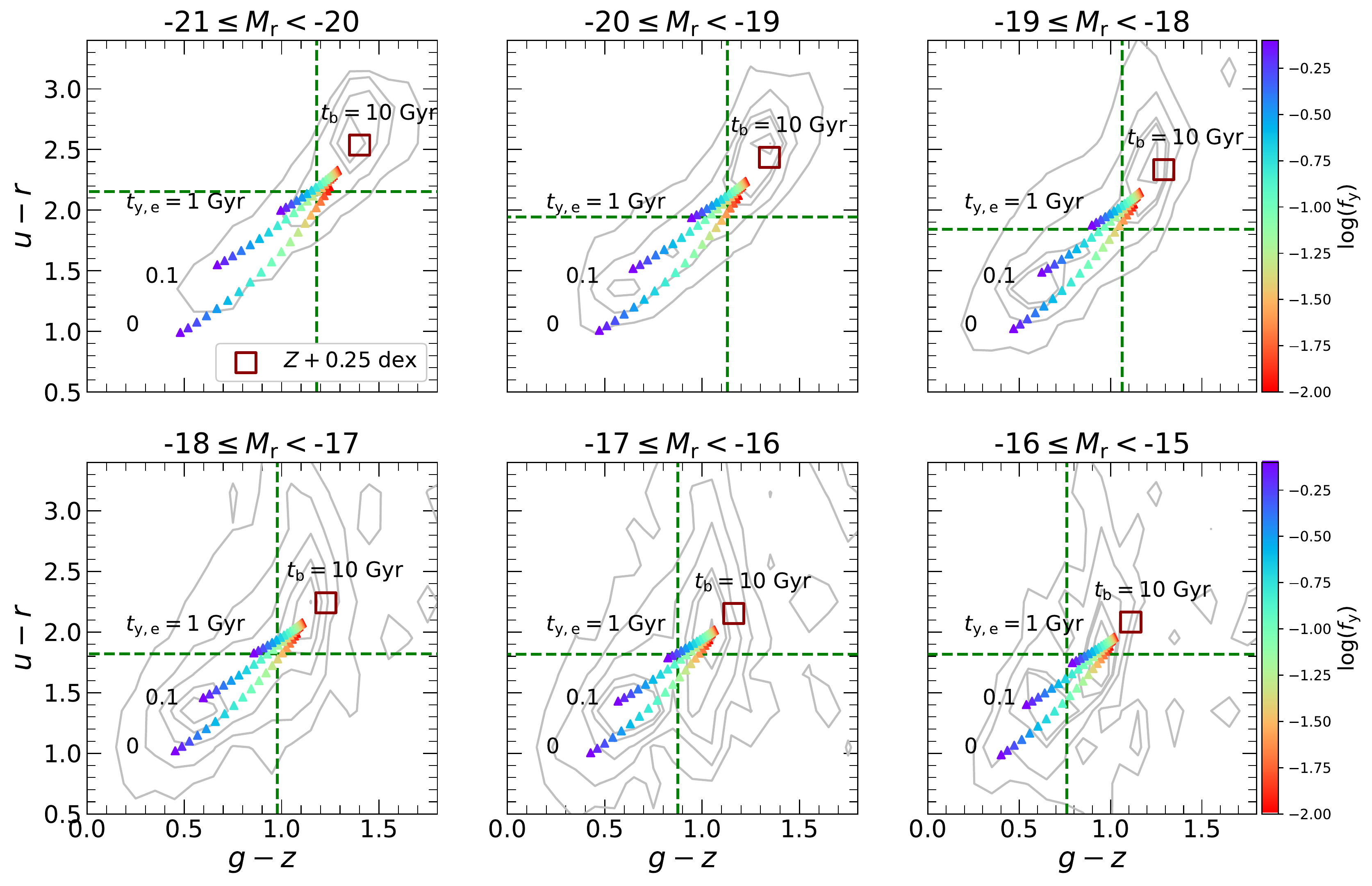}
	\caption{The figures show the results of the two components model, one starburst in early time and a constant star formation rate from some early time to the present day at different absolute magnitude. The age of starburst is $t_{\rm b}=10\,{\rm Gyr}$. The constant star formation starts from $t_{\rm y,o}=5\,{\rm Gyr}$. We plot three cases where $t_{\rm y,e}=0,0.1,1.0\,{\rm Gyr}$. The colors represent the fraction of the young population. The hollow squares represent the results of the starburst model with the metallicity higher than the mean relation of \autoref{eq:MZR} $0.25\,{\rm dex}$. The background grey contours represent the distribution in the groups we measure.}
	\label{fig:old+young_model}
\end{figure*}

Next, we consider a more complex model in which the star formation history
of a galaxy consists of two components, one starburst in early times, and 
a constant star formation rate from some earlier time to the present day.   
Mathematically, the history can be written as follows:
\begin{small}
\begin{equation}
    \Psi\left(t\right)\propto \left(1-f_{\rm y}\right)\delta\left(t-t_{\rm b}\right)+
    \left(\frac{f_{\rm y}}{t_{\rm y,o} -t_{\rm y,e}}\right)
    \mathcal{H}\left(t_{\rm y,o}-t\right)\mathcal{H}(t-t_{\rm y, e})\,,
\end{equation}
\end{small}
where $\mathcal{H}\left(t\right)$ is the step function:
$\mathcal{H}(x)=1$ for $x>0$ and $=0$ otherwise; 
$f_{\rm y}$ is the fraction of the young population; 
$t_{\rm b}$ is the age of the old population;  
$t_{\rm y, o}$ and $t_{\rm y, e}$ are the look-back times of the onset 
and ending of the constant star formation period, respectively.  
For simplicity, we fix the age of the old population at 
$t_{\rm b}= 10\, {\rm Gyr}$, and the onset of the young population at 
$t_{\rm y,o} = 5\,{\rm Gyr}$. We test three cases where 
$t_{\rm y,e}= 0$, $0.1$ and $1.0\,{\rm Gyr}$, respectively. 
We note that our results are not sensitive to 
the exact values chosen for $t_{\rm b}$ and 
$t_{\rm y,o}$.
We use the same method 
as described above for the single burst model to estimate the color-magnitude 
relations for different values of the young fraction, $f_{\rm y}$.  

In \autoref{fig:old+young_model} the triangles show the predicted 
$(g-z)$ versus $(u-r)$ relation on top of contours showing the observed color-color 
relation for group galaxies. The three tracks in each panel represent the 
three different assumptions for the value of $t_{\rm y,e}$, while the color 
scale along each track represents the value of the young fraction, $f_{\rm y}$. 
In each panel, the open square shows the result of a burst at 
$t_{\rm b} = 10\,{\rm Gyr}$, with a metallicity of 0.25\,dex above that assumed 
for the $t_{\rm y,e}$ sequences plotted in the panel. As one can see, the model 
predictions can roughly cover the color-color sequence for galaxies of 
different luminosity in the observational data. In particular, the $f_{\rm y}$-sequence 
assuming $t_{\rm y,e}=0.1\,{\rm Gyr}$ follows well the trend from the red to blue 
modes. Galaxies in the red mode are old in their stellar population, and there is 
little room for them to contain a significant amount of recent star formation. 
The blue mode is consistent with having a constant star formation rate 
continuing to the present day, but more than half of the stellar mass is still 
contained in the old population, even for low-luminosity galaxies. This is consistent with 
results obtained by \citet{Weisz2011} from modeling the color-magnitude relation 
of individual stars in nearby low-mass galaxies, 
by \citet{Lu2014empirical,Lu2015empirical} from 
empirical models of galaxy formation in dark matter halos, 
and by \citet{Zhou2020} from stellar population synthesis modeling of 
the star formation history of low-mass galaxies in the SDSS-IV/MaNGA survey.
As expected, the allowed young fraction depends on the value of $t_{\rm y,e}$.
For $t_{\rm y,e} = 1\,{\rm Gyr}$, the young fraction $f_{\rm y}$ near the 
demarcation line is about 30\% at $M_{\rm r}<-17$. In contrast, if $t_{\rm y,e}=0$, 
the value of $f_{\rm y}$ near both demarcation lines are less than $5\%$. 
For $M_{\rm r}>-17$, the value of $f_{\rm y}$ at the demarcation lines can be a 
factor of two higher. Overall, a large fraction of stars are still in the old 
population, independent of galaxy luminosity. 

For low-luminosity galaxies, the observed $(u-r)$ color 
distribution at given $(g-z)$ has large scatter, which is not well reproduced 
by varying the age and metallicity of the stellar population.  
We note, however, that the $u$-band data from the SDSS imaging survey 
are relatively shallow, with the 95\% completeness limit at 
$u\approx 22.0$ \citep[][]{SDSSDR72009}. Thus the results shown in 
\autoref{fig:gz-ur_dis}, obtained from data at $r < 21$, may be affected 
significantly by the detection limit of the $u$-band, particularly for faint 
red galaxies that are intrinsically faint in ultraviolet. The model presented here  
is also very simple, and ignores any dust attenuation that may have a 
significant impact on the $(u-r)$ color. Furthermore, there is also degeneracy 
between age and metallicity in the spectral synthesis model used here.  
In our model, we only use the mean metallicity-mass relation to make model predictions.  
We find that including the scatter in the metallicity-mass relation 
has the effect of stretching the color-color relation along individual 
$t_{\rm y,e}$ sequence, without broadening it significantly. However, 
there might be correlations between metallicity and other galaxy properties, 
which may change the color-color relation and is not included in our model. 

In conclusion, our results indicate that 
(i) member galaxies of clusters/groups in the red mode are dominated by old stellar populations, 
    independent of galaxy luminosity (mass);
(ii) member galaxies of clusters/groups in the blue mode also contain a large 
     fraction (more than a half) of old stars in their stellar population;
(iii) a more-or-less constant star formation extending to the recent past is needed 
      for galaxies, particularly low-luminosity ones, in the blue mode;
(iv) for galaxies in the green valley [near demarcation lines in the $(u-r)$-$(g-z)$ plane], 
     only a small fraction of their stars can have ages lower than $1\,{\rm Gyr}$, 
     indicating that their star formation was quenched more than $1\,{\rm Gyr}$ ago. 

\subsection{The old fraction and evidence for a characteristic stellar mass scale}
\label{sec:characteristic}

\begin{figure*}
    \centering
	\includegraphics[width=0.75\textwidth]{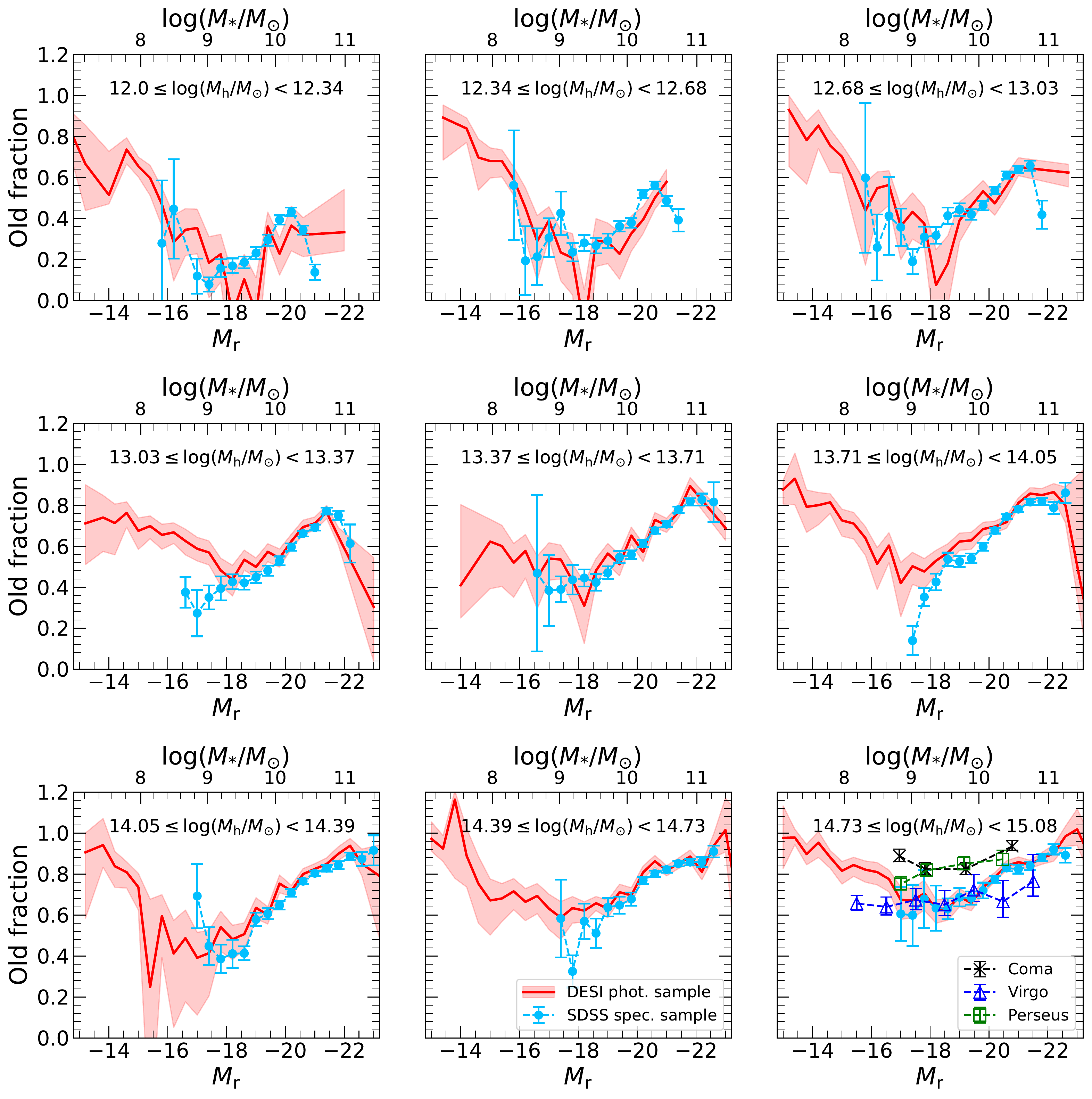}
    \caption{The old fraction defined as the fraction of the satellites in 
    	red mode by the $(g-z)$ color as a function of $r$-band absolute magnitude 
    	and stellar mass, measured for different intervals of halo mass 
    	as indicated. In each panel the red line shows the measurements from 
    	the DESI photometric sample, and the shaded region shows the standard deviation 
    	estimated from 200 bootstrap re-samplings. The light blue dots connected by 
    	dashed lines represent the result from the SDSS spectroscopic sample. 
    	The three additional symbols/lines in the last panel show the red fractions 
    	measured by \citet{Weinmann2011} for the Coma, Virgo and Perseus clusters.}
    \label{fig:fo}
\end{figure*}

In last subsection, we have seen that there is a well-defined population 
of red galaxies that are old. Here we examine the fraction of the old population 
as a function of galaxy luminosity (mass) in halos of different mass.  
The old fraction is calculated from the conditional luminosity function 
of red galaxies divided by that of the total population, and the results are shown in 
\autoref{fig:fo}. The red lines are the results obtained from the DESI-DR9, 
and the shaded regions represent the standard deviation estimated from 200 bootstrap
re-samplings. The average old fraction increases with halo mass, particularly 
in the intermediate luminosity range. There is a valley-like structure of the old 
fraction as a function of galaxy luminosity for all halo mass bins. 
The minimum of the old fraction occurs at $M_{\rm r}\sim -18$, pretty much
independent of halo mass. This result indicates the existence of a 
characteristic luminosity, $M_{\rm r}\sim -18$ (corresponding roughly 
to a stellar mass of $10^{9.5}{\rm M_{\odot}}$), at which the old fraction 
of galaxies is the lowest. 

As a check of our results, we also estimate the old fraction using the 
spectroscopic sample from SDSS, which is much shallower than the DESI-DR9.
The memberships of individual galaxy groups used here are adopted directly
from \citet{Yang2007}. Each galaxy is weighted by $1/(V_{\rm max}\mathcal{C})$ to compensate 
the apparent magnitude limit and redshift incompleteness, where 
$V_{\rm max}$ and $\mathcal{C}$ are both from the NYU-VAGC. The results obtained this 
way are shown as blue solid points in \autoref{fig:fo}, with error bars again obtained from 
200 bootstrap samples. These results are consistent with those obtained from DESI-DR9
over the luminosity range where the old fraction can be measured reliably 
from the spectroscopic sample. For the lowest three halo mass bins where 
the measurements from the spectroscopic sample can go below $M_{\rm r}=-18$, 
an upturn towards the fainter end can also be seen.   
All these indicate that the use of the photometric data does not introduce any 
significant systematic error in our results.  

\subsection{Comparisons with previous observations}
\label{sec:local_group}

\begin{figure*}
    \centering
	\includegraphics[width=0.75\textwidth]{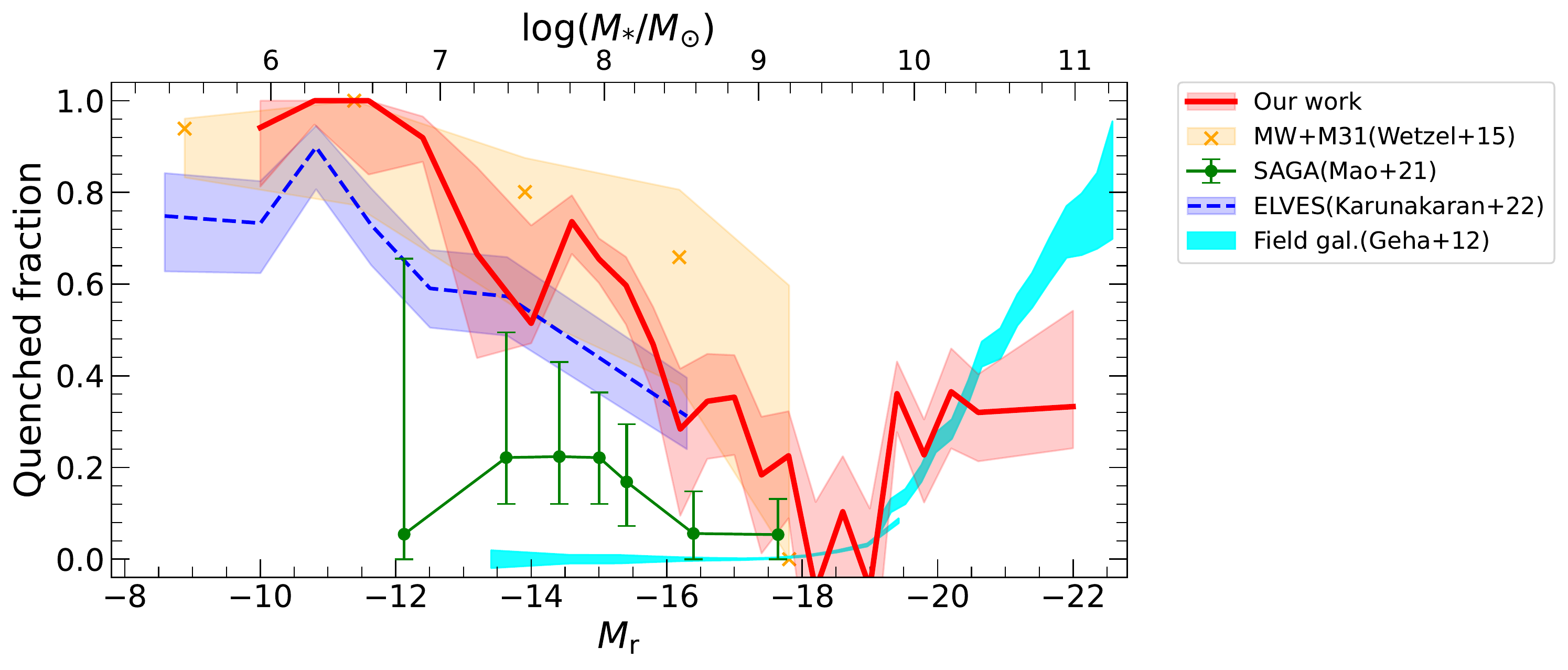}
    \caption{Our measurements of old fraction of satellite galaxies in Milky Way mass halos 
    	with $12\leq \log(M_{\rm h}/M_\odot)<12.34$ (red line) 
    	is compared with the quenched fraction of the satellites in the Milky Way and M31 
    	system (yellow crosses; \citealt{Wetzel2015MW}), as well as 
    	measurements from the ELVES \citep[blue dashed line;][]{Karunakaran2022} and SAGA 
    	\citep[green symbols/line;][]{Mao2021SAGA2} surveys. The result for field 
    	galaxies in SDSS from \citet{Geha2012}  is plotted as the cyan shaded region for comparison.}
    \label{fig:fq_MWlike}
\end{figure*}

There have been many studies to measure the fraction of quenched/red 
galaxies as functions of galaxy properties and/or dark matter halo mass. 
In most cases these studies are based on spectroscopic samples limited to 
relatively bright galaxies, while other studies are focused on individual 
systems in the nearby Universe. In the bottom-right panel of \autoref{fig:fo}
we show the red fractions measured by \citet{Weinmann2011} for the Coma, Virgo, and Perseus 
clusters. Similar to our result, the red fraction in Coma cluster shows a dip at around 
$M_{\rm r}\sim-18.5$, although the overall amplitude is higher.  
The red fractions in the Virgo cluster are comparable to our result, 
but with a weaker dependence on luminosity and no obvious dip at any 
luminosity. The Perseus cluster also shows no significant dip at the 
intermediate luminosity, and red fractions are comparable to 
those of the Coma cluster and higher than our result. 
We note that the red/blue galaxy populations in \citet{Weinmann2011}
were identified using the $(g-r)$ color, which might  
mis-classify some of the faint galaxies that are red in $(g-z)$ but 
blue in $(g-r)$. The higher red fractions in Coma and 
Perseus may be reflecting the overall increase of the red fraction 
with increasing halo mass, given that the halo masses of 
the two clusters, $M_{\rm h}=2.7\times10^{15}M_\odot$ for Coma 
\citep{ComaCluster2007} and $M_{\rm h}=2.2\times10^{15}M_\odot$ 
for Perseus \citep{PerseusCluster2020}, are higher than both the average mass 
of groups in that panel, $M_{\rm h}=7\times10^{14}M_\odot$, and 
the mass of te Virgo cluster, $M_{\rm h}=6.3\times10^{14}M_\odot$ 
\citep{VirgoCluster2020}. Given these differences, the comparison with 
our results is not straightforward. 

Measurements of quenched fractions for faint satellite galaxies 
have also been obtained recently for nearby systems 
similar to the Local Group. In \autoref{fig:fq_MWlike} we show the 
old fraction of galaxies as a function of $M_{\rm r}$ obtained by us for 
Milky Way-like halos with $12.0\leq(M_{\rm{h}}/M_\odot)<12.34$, 
in comparison with the quenched fractions obtained by \citet{Wetzel2015MW} for 
the MW$+$M31 system, by \citet{Geha2017SAGA1} and \citet{Mao2021SAGA2}
using the SAGA survey of 127 satellite galaxies around 36 Milky Way (MW) 
analogs at $z\sim 0.01$, and by \citet{Carlsten2022} 
and \citet{Karunakaran2022} using the ELVES survey of a nearly volume-limited 
sample of dwarf satellites down to $M_{\rm V}\sim -9$. Our result is in good agreement 
with the result of both the MW$+$M31 halo and the ELVES survey. The quenched fractions 
from the SAGA survey also show an upturn as the luminosity decreases,
but overall they are significantly lower than both the MW+M31 result and
our measurement. \citet{Karunakaran2022} made careful comparisons 
between the SAGA and ELVES results, and attributed the lower quenched 
fractions from SAGA to sample incompleteness at the low-mass end 
caused by the adopted absolute magnitude cut for satellites ($M_{\rm r}<-12.3$).
Note, however, the old fraction in the present paper is defined by the $(g-z)$ color, 
which could be different from the quenched fraction defined in 
other studies using star formation rates. 
As mentioned in \autoref{sec:old_sequence}, 
the old sequence based on the $(g-z)$ color may contain some recent star 
formation, and the true quenched fraction might be lower than the old 
fraction defined in our paper. For comparison, the quenched fraction of 
{\em field} galaxies estimated by \citet{Geha2012} based on the NASA-Sloan 
Atlas catalog \citep[][]{Blanton2011NSA}
is plotted as the cyan shaded region. Different from the member 
galaxies of groups/clusters, galaxies with $M_{\rm r}>-18$
in the field are barely quenched. This again indicates that the upturn 
of the old (quenched) fractions at the faint end is a unique behavior of 
group/cluster galaxies.

\subsection{Comparisons with theoretical predictions}

\begin{figure*}
    \centering
	\includegraphics[width=0.75\textwidth]{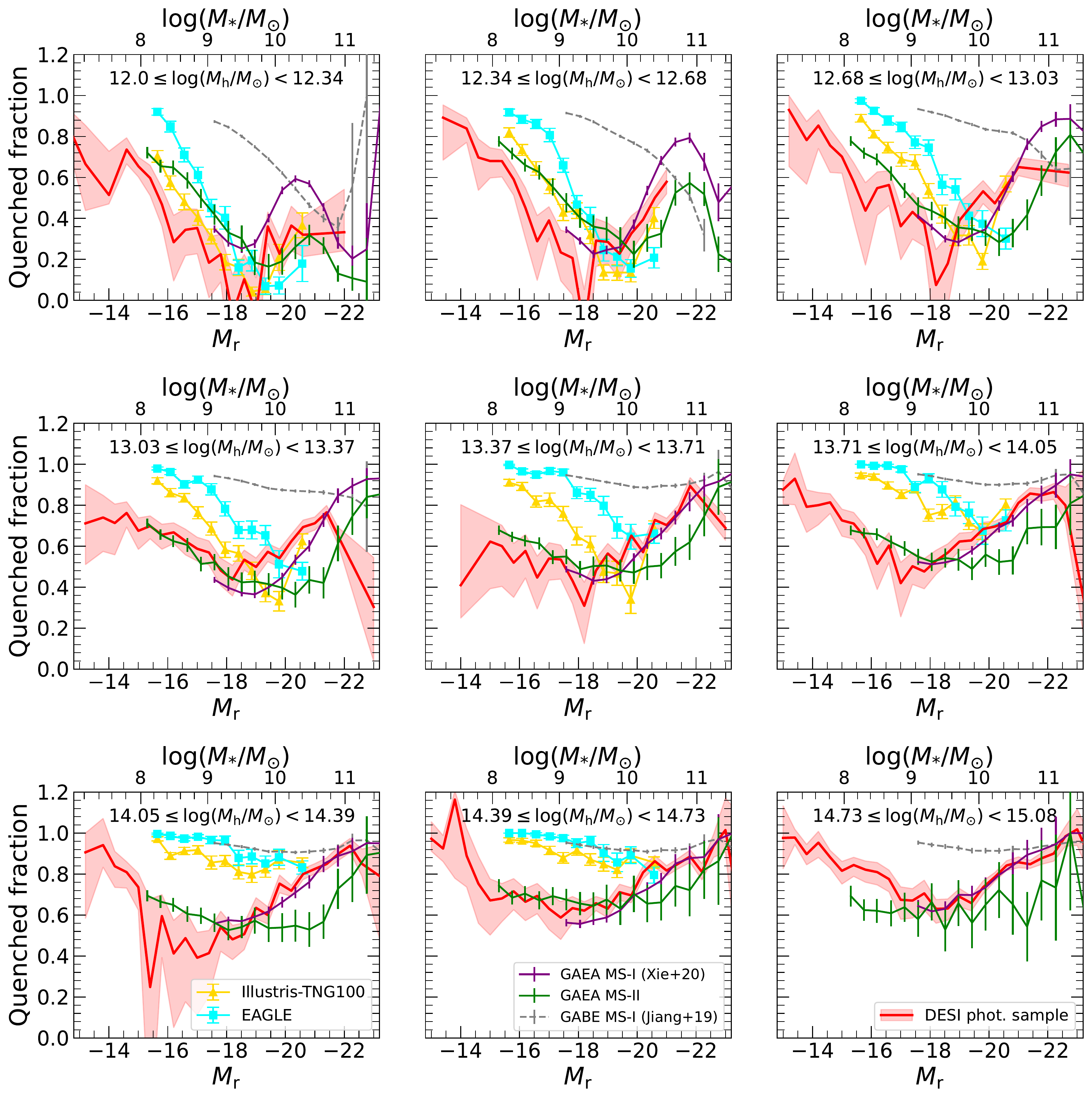}
	\caption{Our measurements of the old fraction in different halo mass 
	intervals (indicated in each panel) are compared to the quenched fractions 
	predicted by the GAEA semi-analytic model of \citet[][purple line for result with
   the Millennium simulation and green line for that with the Millennium-II 
   simulation]{Xie2020}, the GABE 
   semi-analytic model of \citet[][grey dashed line]{Jiang2019GABE}, and the 
   TNG (yellow triangles) and EAGLE (cyan squares) simulations.}
	\label{fig:models}
\end{figure*}

It is also interesting to compare our measurements of the old fraction
with theoretical models which predict the fraction of 
quenched or quiescent satellites for halos of different mass.
In \autoref{fig:models}, we compare our measurements with the predictions 
by two ``state-of-the-art''  hydro-dynamic simulations: 
Illustris-TNG100 \citep[][]{Nelson2019TNG,Pillepich2018TNG} and 
EAGLE \citep[][]{Schaye2015EAGLE,Crain2015EAGLE,McAlpine2016EAGLE}.
For massive galaxies with $M_\ast\ga 10^{10}M_\odot$, both 
simulations can roughly match the observational results, consistent 
with what was found in \citet{Xie2020}. Both simulations can also roughly 
reproduce the Valley-like shape of the quenched fraction, although the 
minimum of the quenched fraction in the simulations occurs 
at a slightly higher luminosity ($M_{\rm r}=-19\sim-20$)
than in the observation ($M_{\rm r}\sim-18$). Quantitatively, 
Illustris-TNG100 can match the observation down to $M_\ast\sim10^8M_\odot$
for halos with mass $M_{\rm h}\lesssim 5\times10^{13}M_\odot$, while 
EAGLE overpredicts the quenched fractions for galaxies below the 
characteristic mass for halos of all masses except for Milky-Way mass
halos. For massive halos above $M_{\rm h}\sim 5\times10^{13}M_\odot$, 
both simulations significantly over-predict the quenched fraction for 
satellites with $M_\ast<10^{10-10.5}M_\odot$. 
Our result indicates that quenching mechanisms
implemented in current hydrodynamic simulations are able to capture 
the evolutionary processes for satellites in halos with low-to-intermediate 
mass and massive satellites in massive halos, but not for  
low-mass satellites in massive halos. 

In the same figure, we also compare our result with the quenched fractions 
in two different semi-analytic models of galaxy formation: 
Galaxy Assembly with Binary Evolution \citep[GABE;][]{Jiang2019GABE},
and GAlaxy Evolution and Assembly \citep[GAEA;][]{GAEA2016,Xie2017,DeLucia2019,Xie2020}.
The GABE model follows the same recipies of satellite quenching 
as in early models \citep[e.g.][]{Guo2011SAM}, where the quenching 
of satellites occurs only by {\em starvation} or equivalently 
{\em strangulation} \citep{Larson1980,Balogh2000b}, with a gradual removal instead 
of instantaneous stripping of hot gas after they fall into their host halos. 
Consistent with what is previously known 
\citep[e.g.][]{Font2008,Guo2011SAM,Hirschmann2014}, 
\autoref{fig:models} shows that this model predicts too high a quenched 
fraction at 
all luminosities except at the bright end ($M_{\rm r}\lesssim -21$, 
or $M_\ast\ga 10^{11}M_\odot$), and this is true for all halo masses. 

\citet{Xie2020} compared the fraction of 
quenched galaxies as a function of galaxy stellar mass down to 
$M_\ast\sim 10^9M_\odot$ obtained from the SDSS by \citet{Wetzel2012} 
and \citet{Hirschmann2014} with those predicted by four semi-analytic models: 
L-Galaxies  \citep{L-Galaxies2017}, SAG \citep{SAG2018},
GAEA \citep{Xie2017}, and a modified version of the GAEA model by the 
authors themselves with improved treatments on environmental effects.
All the models were found to reproduce the overall trends in the 
observational data (see Fig.9 of \citealt{Xie2020}). In \autoref{fig:models}, 
the purple lines with (Poisson) error bars repeat the results of the 
modified GAEA code of \citet{Xie2020} applied to the Millenniuum Simulation
\citep[MS;][]{Springel2005MS}. Here we extend the predictions to a lower 
mass limit ($M_\ast\sim10^8M_\odot$) by applying the same code to the 
Milleniuum-II Simulation \citep[MS-II;][]{Boylan-Kolchin2009MSII}.  
As one can see, the GAEA model performs much better than GABE
in matching our measurements at all halo masses, presumably because 
of the better treatment for satellite quenching. The differences in the 
model prediction between MS and MS-II are produced by the difference
in numerical resolution, and by the fact that the model parameters 
were tuned only according to the MS. 
As pointed out in \citet{DeLucia2019} and \citet{GAEA2016}, 
the success of GAEA is achieved not only by the improved 
treatment of satellite quenching (e.g. environmental effects 
considered in \citealt{Xie2020}), but more importantly by 
the large cold gas mass in satellites at the time of infall
due to mechanisms implemented to regulate the gas-to-star 
conversion prior to infall, and by a lower rate implemented 
for gas heating by supernova explosions and stellar winds 
after the infall. We note that the same success is also achieved by 
the updated L-Galaxies model of \citet{L-Galaxies2017}
by using a more comprehensive treatment of satellite quenching.

As pointed out above, the old populations selected by $(g-z)$ in 
our case may still contain recent star formation, and they 
could be classified as star-forming if the quenched fraction 
is defined according to star formation rate. Therefore, the 
comparison results presented here should be taken with caution. 
A detailed comparison of the different models/simulations with our 
measurements, with the different definitions of quenched fraction
being carefully taken into account, is needed to understand how 
our results can be used to constrain quenching processes quantitatively.
This is beyond the scope of the present paper and 
we will come back to it in a forthcoming paper. 

\section{Discussion}
\label{sec:discussion}

\subsection{Implications of the characteristic mass}

It is interesting to see how the characteristic luminosity at $M_{\rm r}\sim -18$ 
($M_\ast\sim10^{9.5}M_\odot$) found here for member galaxies 
of clusters/groups compares to other observations along the same line. 
Traditionally, a $B$-band magnitude at $-18$ has been used to 
separate elliptical galaxies into dwarf and normal populations. 
These two populations seem to follow two distinct trends in the 
relation between surface brightness and absolute magnitude, with 
the dwarf population showing a much steeper increase of surface brightness 
with luminosity \citep[][]{Kormendy1977,Kormendy1985,Graham2003,Ferrarese2006,Kormendy2009}. 
Using a large sample of SDSS galaxies, \citet{Shen2003}
found  that red dwarf galaxies with $-16<M_{\rm r}< -18$
have similar sizes, while brighter red galaxies
show a strong increase in size with luminosity (see their fig.13). 
These previous results suggest that there may be a dichotomy 
in structural properties between galaxies separated at 
$M_{\rm B}\sim-18$. It is more meaningful to make 
the comparison in terms of stellar mass instead of luminosity in 
different bands. According to the empirical stellar mass 
estimators in \citet{Bell2001} and \citet{Bell2003}, $M_{\rm B}=-18$ 
corresponds to a stellar mass $M_\ast=10^{9.1}\sim10^{9.9}M_\odot$
for a typical color range of $B-R=0.8\sim1.5$, consistent
with the average mass of $M_\ast\sim10^{9.5}M_\odot$
for galaxies with $M_{\rm r}\sim-18$ in our sample.
This implies that, for galaxies in groups/clusters, the dichotomy 
in the stellar population we find here and that in galaxy structure found 
in previous studies may be driven by some common processes 
that operate in groups/clusters of galaxies. For example, tidal stripping
and galaxy interaction in dark matter halos may not only affect gas contents 
of satellite galaxies, but also have impacts on the stellar content 
and structure of galaxies. These effects are expected to be more effective for 
lower-mass (fainter) galaxies with lower surface densities 
\citep[e.g.][]{Li2012,Zhang2013}. Analyses based on galaxy groups/clusters 
suggest that strangulation may be more important for quenching star formation 
in satellites more massive than $M_\ast\sim10^{9.5}M_\odot$ 
\citep[e.g.][]{Weinmann2006a,Weinmann2006b,vandenBosch2008,vanderWel2010,vonderLinden2010}.
In the strangulation process, the hot gas halo around a galaxy is removed 
gradually, and so the quenching time scale associated with it is expected to be
longer than that caused by the more violent processes, such as 
ram-pressure and tidal stripping. In a more recent study by 
\cite{LiPengfei2020} based on the SDSS groups, the quenched fraction of 
member galaxies of groups is found to exhibit a minimum, but at a 
characteristic mass scale that depends on both the central galaxy mass and the 
host halo mass. Quantitatively, their characteristic mass is comparable to ours 
only for Milky-Way size halos, i.e. at $M_\ast\sim10^{9.5}M_\odot$; 
as halo mass increases, the stellar mass for the demarcation increases 
up to $M_\ast>10^{10.5}M_\odot$ for the most massive halos with 
$M_{\rm h}>10^{14}M_\odot$ (see their fig.3). 
As pointed out by the authors, their results imply that galaxy 
quenching is not simply driven by the central-satellite dichotomy,
but determined by the interaction between internal and environmental 
processes. Our results for faint satellites at $M_\ast<10^{9.5}M_\odot$
combined with those obtained earlier for more massive galaxies seem 
to indicate that there are two characteristic mass scales, one is 
a fixed mass scale $M_1\sim10^{9.5}M_\odot$ found here, and the other one, $M_2$, 
which depends on halo mass as found in \cite{LiPengfei2020}.
For red galaxies at $M_\ast<M_1$ (which are predominately satellites), 
violent processes such as tidal and ram-pressure stripping 
may play the dominant role in quenching star formation, while for galaxies above $M_1$, 
the quenching may be driven by a gradual removal of their hot halo gas 
through strangulation if $M_\ast<M_2$ or by internal processes, 
such as AGN feedback, if $M_\ast>M_2$. 

\subsection{The faint-end upturn and implications 
for galaxy formation at high redshift}

The faint-end upturn in the luminosity function of galaxies in groups/clusters 
has long been in debate, as mentioned in Introduction. 
In the present paper, we have obtained the most reliable measurements 
of CLFs for groups with halo mass $M_{\rm h}\ga 10^{12}M_\odot$ and
for galaxies down to $M_r=-10\sim-12$, using a well-defined sample of central 
galaxies with halo mass information from the groups 
of \citet{Yang2007}, the updated photometric sample of DESI, and the use of 
the $(g-z)$ color that is able to provide a more reliable separation between the old and young 
galaxy populations. Our results confirm the presence of a faint-end 
upturn for the red/old satellite population in halos of a wide range of mass, 
as shown in \autoref{fig:clf_red}. For the total CLFs, although the faint-end slope we obtain here, 
$\alpha\sim-1.6$, is somewhat steeper than that obtained earlier, it is actually 
comparable to that for field galaxies. For example, \citet{Blanton2005} obtained $\alpha\sim-1.5$ after 
correcting for the incompleteness in the SDSS sample, and \citet{Li2022} obtained 
$\alpha\sim-1.6$ (see their figure 8) after correcting the cosmic variance effect 
in the SDSS sample \citep{Chen2019}. 
Recent measurements of galaxy stellar mass function (GSMF) 
in the GAMA sample showed that the faint-end slope is about 
$\alpha\sim -1.53$ \citep[][]{Driver2022}, and that there is an upturn in the GSMF  
at $M_\ast\sim10^{9.5}M_\odot$. Such an upturn may be produced by that seen 
in the CLF, particularly of red satellites. However, the GSMF obtained 
from the SDSS sample by \citet{Chen2019} shows an upturn mass 
that is slightly larger, at $M_\ast\sim10^{10}M_\odot$.   
Since the upturn in the total GSMF of field galaxies is relatively 
weak, the feature may be more difficult to quantify.
The faint-end upturn in the CLF is seen only for 
red/old satellites, with a slope of $\alpha\sim-1.8$ and a turnaround luminosity 
at $M_{\rm r}\sim-18$, both of which are almost independent of halo mass 
(see \autoref{fig:para_color}). The CLFs of blue satellites can be well described 
by a single Schechter function with a flatter slope of $\alpha\sim-1.25$, 
independent of halo mass. As shown by \autoref{fig:clf_blue} and \autoref{fig:clf_red}, 
it is the blue population that raises the amplitude of the luminosity function (LF) at 
$M_{\rm r}<-18$, producing a total LF without a significant upturn at the faint end, 
although the red/old population dominates the total abundance at 
the faint end. In the highest halo mass bin, which corresponds to the mass range 
of nearby rich clusters used in earlier studies, the total LF is dominated by 
the red population at almost all luminosities. This explains why the 
faint-end upturn was observed for those rich clusters in earlier studies. 
Controversial results in previous studies may be caused, at least 
partly, by the different fractions of the red/old population due to different 
photometric bands used to select satellite galaxies. 

Although the origin of the faint-end slope of the total CLF, and in particular 
the upturn observed for red galaxies, are still poorly understood,
it is clear that they contain important information about the formation 
and evolution of low-mass galaxies. For instance, in the empirical model 
developed by \citet{Lu2014empirical} and \citet{Lu2015empirical}, 
a star formation burst at $z>2$ was required for dwarf galaxies with 
$M_\ast\lesssim 10^{9}M_\odot$ in order for their model to reproduce 
the faint-end upturn in the CLF of clusters obtained by \citet{Popesso2006}. 
A natural prediction of this model is that the predicted mean stellar age of 
galaxies has a minimum at $\sim10^{9.5}M_\odot$ (see their fig.14), 
which is consistent with our finding of the characteristic mass 
in the old fraction as a function of galaxy luminosity. 
Another prediction from the model of \citet{Lu2014empirical} and \citet{Lu2015empirical}
is steep faint-end slopes of galaxy LFs at $z>2$, 
which implies higher stellar mass (or luminosity) to dark matter mass ratios 
in low-mass halos as one goes to higher redshift (see fig.3 of \citealt{Lu2015empirical}). 
The relation between galaxy luminosity/stellar mass and host halo mass has often been 
obtained from the abundance matching model, by matching the abundance of galaxies 
above a luminosity threshold with the abundance of subhalos with mass 
(defined at the time when the subhalo was accreted by its host halo) 
above a threshold. \citetalias{Lan2016} extended the abundance matching 
model to relate the faint-end slope of CLFs ($\alpha$) with the ``efficiency'' of star 
formation in low-mass halos using a parameterized form $L\propto m^{\beta}$,
where $L$ is galaxy luminosity and $m$ is the un-evolved subhalo mass.
Assuming that the number density is dominated by faint galaxies, i.e. 
$\alpha<-1$, this matching leads to a simple relation between $\alpha$ and $\beta$: 
$\beta\approx -0.8/(1+\alpha)$. 
For blue satellites with $\alpha=-1.25$ from our CLFs, this relation
implies that $L\propto m^{3}$, which is steeper than the relation of 
$L\propto m^{3/2}$ obtained by \citetalias{Lan2016}. The difference 
is caused by the significant faint end upturn in the CLFs of blue satellites 
obtained in \citetalias{Lan2016}  
(see the comparison of CLFs in \autoref{fig:clf_blue}).
As demonstrated in~\autoref{sec:comparison}, a fraction of the blue galaxies 
identified by the $(u-r)$ color in \citetalias{Lan2016} are actually red (and old) 
when classified according to $(g-z)$, which may explain the discrepancy.
For the red population, both \citetalias{Lan2016} and our results 
give a faint-end slope of $\alpha\approx-1.8$, implying that $L\propto m$. 

As suggested in \citetalias{Lan2016}, the different faint-end slopes, 
or equivalently the different $L-m$ relations, for the red and blue populations
imply a dichotomy in the formation processes of the present-day galaxy 
population, with a characteristic redshift $z_{\rm c}$ at which the mode of 
star formation made a change. At $z>z_{\rm c}$, the star formation efficiency 
(as measured by the stellar mass to halo mass ratio) in low-mass halos is 
roughly independent of halo mass, while at $z<z_{\rm c}$ the efficiency 
increases rapidly with halo mass. For satellite galaxies, their star 
formation continues until they became satellites in larger systems.
Thereafter the star formation might be quenched and their stellar mass 
would not change much. Thus, for the old population of satellite
galaxies which formed most of their stars early, their stellar mass 
distribution should inherit that at the time of formation, hence the steep 
slope. In contrast, for the young population of satellites, which became 
satellites late and formed most of their stars in a mode similar to that 
of field galaxies, their stellar mass distribution should resemble that 
of present-day field galaxies, as is seen in our results. This scenario 
is similar to that expected from the pre-heating model of 
\citet{MoMao2002,MoMao2004}, where the transition from an early phase 
of efficient star formation to a later phase of reduced star formation 
efficiency is predicted to occur at a time when the intergalactic gas
was heated significantly.  

 Environmental effects after a galaxy falls into its host halo
may also produce the faint-end upturn in the red/old population, 
if these effects are to convert a fraction of satellites from 
the young population to the old in a systematic way. Quantitatively, 
to obtain the faint-end slope of $-1.8$ for the old population from 
a slope of $-1.6$ for the total population, the conversion fraction 
should scale with luminosity as $\propto L^{-0.2}$. Thus, 
the observed faint-end upturn may also be explained by environmental 
effects (nurture), instead of formation processes (nature). However, 
it is unclear how the vastly different environments, as represented 
by the large range of halo mass, can produce the same trend in the 
conversion and a single characteristic mass scale at 
$\sim 10^{9.5}{\rm M}_\odot$.

This `nature and nurture' degeneracy can be broken by observing the luminosity/stellar 
mass functions of low-mass galaxies at high $z$. If the steep slope 
originates from `nature' (formation), then high-$z$ galaxies, which are 
the progenitors of old galaxies observed in present-day clusters/groups, 
should be expected to have a steep luminosity/stellar mass function at the 
faint end. On the other hand, if the steep slope is produced by `nurture'
(later environmental effects), then a steep slope is not expected 
for galaxies at high redshift.
Observations of galaxy luminosity/stellar mass functions at high $z$ 
down to a stellar mass below $\sim 10^{9.5}M_\odot$ are crucial to test 
the different scenarios. Recent observations based on ALMA 
\citep[e.g.][]{Gruppioni2020,Loiacono2021,Khusanova2021} found a much 
higher cosmic star formation rate density 
at $4 \lesssim z\lesssim 6$ than previously derived from UV-selected galaxies, 
suggesting that star formation may be more efficient at high $z$. 
More recent results from the Early Release Observations and Early Release 
Science Program of the JWST \citep[e.g.][]{Donnan2022,Harikane2022b} also 
revealed a high cosmic density of star formation rate at high $z$.  
Clearly, with the advent of such observational data, we should be able  
to distinguish these different scenarios.  


\section{Summary}
\label{sec:summary}

In this paper, we use the DESI-DR9 imaging data down to the $r$-band magnitude limit 
of $r=23$ and the group catalogue of galaxies from the SDSS spectroscopic survey 
to study the galaxy populations in groups/clusters with halo mass $M_{\rm h}\ge 10^{12}M_{\odot}$ 
and redshift in the redshift range $0.01\le z\le 0.08$. With central galaxies from the group catalog and 
satellites galaxies sampled with the imaging data, we measure the conditional luminosity 
functions (CLFs) in groups of different halo mass down to $M_{\rm r}=-10\sim-12$ mag, 
more than 2 magnitudes fainter than the limits achieved in previous investigations. 
We use the $(g-z)$ color to separate satellite galaxies into red and blue populations,
and measure the CLFs separately for both. We calculate the fraction of the red population as a 
function of galaxy luminosity in halos of different mass, and use stellar population 
synthesis models to demonstrate the color bimodality of group galaxies is driven 
mainly by the stellar age of galaxies combined with the observed metallicity-mass 
relation. Our main results can be summarized as follows.
\begin{enumerate}
    \item The CLFs we obtain are in good agreement with previous measurements,
    but with better statistics and down to fainter luminosities thanks to the 
    much deeper imaging data from DESI-DR9. 
    \item We find clear bimodal distributions of group galaxies in $(g-z)$ color, 
    which can be used to separate galaxies into red and blue modes 
    that are cleaner than those from the widely-used $(u-r)$ color, particularly for faint galaxies. 
    \item Our CLF measurements of red and blue galaxies are broadly consistent 
    with those obtained previously. We also find a clear upturn in the faint-end of the  
    CLFs for red galaxies, with a faint-end slope $\alpha\approx -1.8$, similar to the slope of the 
    halo mass function at the low-mass end predicted by the current $\Lambda$CDM model. 
    \item Our results using stellar population synthesis models indicate that group galaxies in the red 
    mode are dominated by old stellar populations, independent of galaxy luminosity (mass),
    although the red mode becomes bluer as luminosity decreases due to the change of metallicity 
    with luminosity.
    Group galaxies in the blue mode also contain a large 
    fraction of old stellar populations, but a more-or-less constant star formation 
    extending to the recent past is needed to match their colors, particularly 
    for low-luminosity galaxies. 
    \item The fraction of old galaxies as a function of galaxy luminosity has a minimum 
    at a characteristic luminosity $M_{\rm r}\sim -18$ independent of halo mass,
    which corresponds to a stellar mass scale of $M_\ast\sim10^{9.5}M_\odot$. 
    This luminosity/mass scale is comparable to the characteristic 
    luminosity with which galaxies show a dichotomy in surface brightness.
    This suggests that the dichotomy in the old fraction found here for group/cluster 
    galaxies and the dichotomy in galaxy structure found earlier may be driven 
    by similar processes.
    \item The rising of the old fraction at the faint end is consistent with the quenched 
    fraction obtained for the Milky Way and M31 system and from the ELVES survey. 
    \item Current semi-analytic models of galaxy formation are able to reproduce the 
    quenched fraction down to $M_\ast\sim10^{8}M_\odot$, 
    while both the Illustris-TNG and EAGLE simulations over-predict the quenched fraction 
    of low-mass satellites ($M_\ast\lesssim 10^{10}M_\odot$)
    in halos more massive than the Milky Way halo mass.
    \item Our results provide important information about the formation and quenching of satellite galaxies.
    The observed upturn in the number of old galaxies in the faint end, which suggests that the stellar masses 
    in these galaxies are proportional to their halo masses, indicates that star formation 
    and feedback efficiencies are independent of halo mass during the formation of these galaxies. 
    This implies that the faint-end slope of the luminosity/stellar mass function of high-$z$ 
    galaxies is steep with $\alpha\sim -1.8$, a prediction that can be tested by JWST observations of  
    low-mass galaxies ($M_*<10^{9.5}{\rm M}_\odot$) at high $z$. The presence of a well-defined 
    characteristic mass at $M_*\sim 10^{9.5}{\rm M}_\odot$ in the old population fraction of satellites 
    in halos of all masses indicates that the quenching of star formation is closely related to the 
    formation and properties of galaxies, rather than their current environments.
\end{enumerate}

\section*{Acknowledgments}
This work is supported by the National Key R\&D Program of China
(grant No. 2018YFA0404502), and the National Science 
Foundation of China (grant Nos. 11821303, 11733002, 11973030, 
11673015, 11733004, 11761131004, 11761141012). 
The authors
acknowledge the Tsinghua Astrophysics High-Performance Computing platform at Tsinghua University for providing computational and
data storage resources that have contributed to the research results
reported within this paper.

Funding for the SDSS and SDSS-II has been provided by the Alfred P. Sloan Foundation, the Participating Institutions, the National Science Foundation, the U.S. Department of Energy, the National Aeronautics and Space Administration, the Japanese Monbukagakusho, the Max Planck Society, and the Higher Education Funding Council for England. The SDSS Web Site is http://www.sdss.org/.

The SDSS is managed by the Astrophysical Research Consortium for the Participating Institutions. The Participating Institutions are the American Museum of Natural History, Astrophysical Institute Potsdam, University of Basel, University of Cambridge, Case Western Reserve University, University of Chicago, Drexel University, Fermilab, the Institute for Advanced Study, the Japan Participation Group, Johns Hopkins University, the Joint Institute for Nuclear Astrophysics, the Kavli Institute for Particle Astrophysics and Cosmology, the Korean Scientist Group, the Chinese Academy of Sciences (LAMOST), Los Alamos National Laboratory, the Max-Planck-Institute for Astronomy (MPIA), the Max-Planck-Institute for Astrophysics (MPA), New Mexico State University, Ohio State University, University of Pittsburgh, University of Portsmouth, Princeton University, the United States Naval Observatory, and the University of Washington.

The Legacy Surveys consist of three individual and complementary projects: the Dark Energy Camera Legacy Survey (DECaLS; Proposal ID \#2014B-0404; PIs: David Schlegel and Arjun Dey), the Beijing-Arizona Sky Survey (BASS; NOAO Prop. ID \#2015A-0801; PIs: Zhou Xu and Xiaohui Fan), and the Mayall z-band Legacy Survey (MzLS; Prop. ID \#2016A-0453; PI: Arjun Dey). DECaLS, BASS and MzLS together include data obtained, respectively, at the Blanco telescope, Cerro Tololo Inter-American Observatory, NSF's NOIRLab; the Bok telescope, Steward Observatory, University of Arizona; and the Mayall telescope, Kitt Peak National Observatory, NOIRLab. Pipeline processing and analyses of the data were supported by NOIRLab and the Lawrence Berkeley National Laboratory (LBNL). The Legacy Surveys project is honored to be permitted to conduct astronomical research on Iolkam Du'ag (Kitt Peak), a mountain with particular significance to the Tohono O'odham Nation.

NOIRLab is operated by the Association of Universities for Research in Astronomy (AURA) under a cooperative agreement with the National Science Foundation. LBNL is managed by the Regents of the University of California under contract to the U.S. Department of Energy.

This project used data obtained with the Dark Energy Camera (DECam), which was constructed by the Dark Energy Survey (DES) collaboration. Funding for the DES Projects has been provided by the U.S. Department of Energy, the U.S. National Science Foundation, the Ministry of Science and Education of Spain, the Science and Technology Facilities Council of the United Kingdom, the Higher Education Funding Council for England, the National Center for Supercomputing Applications at the University of Illinois at Urbana-Champaign, the Kavli Institute of Cosmological Physics at the University of Chicago, Center for Cosmology and Astro-Particle Physics at the Ohio State University, the Mitchell Institute for Fundamental Physics and Astronomy at Texas A\&M University, Financiadora de Estudos e Projetos, Fundacao Carlos Chagas Filho de Amparo, Financiadora de Estudos e Projetos, Fundacao Carlos Chagas Filho de Amparo a Pesquisa do Estado do Rio de Janeiro, Conselho Nacional de Desenvolvimento Cientifico e Tecnologico and the Ministerio da Ciencia, Tecnologia e Inovacao, the Deutsche Forschungsgemeinschaft and the Collaborating Institutions in the Dark Energy Survey. The Collaborating Institutions are Argonne National Laboratory, the University of California at Santa Cruz, the University of Cambridge, Centro de Investigaciones Energeticas, Medioambientales y Tecnologicas-Madrid, the University of Chicago, University College London, the DES-Brazil Consortium, the University of Edinburgh, the Eidgenossische Technische Hochschule (ETH) Zurich, Fermi National Accelerator Laboratory, the University of Illinois at Urbana-Champaign, the Institut de Ciencies de l'Espai (IEEC/CSIC), the Institut de Fisica d'Altes Energies, Lawrence Berkeley National Laboratory, the Ludwig Maximilians Universitat Munchen and the associated Excellence Cluster Universe, the University of Michigan, NSF's NOIRLab, the University of Nottingham, the Ohio State University, the University of Pennsylvania, the University of Portsmouth, SLAC National Accelerator Laboratory, Stanford University, the University of Sussex, and Texas A\&M University.

BASS is a key project of the Telescope Access Program (TAP), which has been funded by the National Astronomical Observatories of China, the Chinese Academy of Sciences (the Strategic Priority Research Program ``The Emergence of Cosmological Structures'' Grant \# XDB09000000), and the Special Fund for Astronomy from the Ministry of Finance. The BASS is also supported by the External Cooperation Program of Chinese Academy of Sciences (Grant \# 114A11KYSB20160057), and Chinese National Natural Science Foundation (Grant \# 12120101003, \# 11433005).

The Legacy Survey team makes use of data products from the Near-Earth Object Wide-field Infrared Survey Explorer (NEOWISE), which is a project of the Jet Propulsion Laboratory/California Institute of Technology. NEOWISE is funded by the National Aeronautics and Space Administration.

The Legacy Surveys imaging of the DESI footprint is supported by the Director, Office of Science, Office of High Energy Physics of the U.S. Department of Energy under Contract No. DE-AC02-05CH1123, by the National Energy Research Scientific Computing Center, a DOE Office of Science User Facility under the same contract; and by the U.S. National Science Foundation, Division of Astronomical Sciences under Contract No. AST-0950945 to NOAO.

%



\software{Astropy \citep{2013A&A...558A..33A,2018AJ....156..123A},  
          NumPy \citep{numpy2020}, 
          SciPy \citep{scipy2020},
          Matplotlib \citep{Matplotlib2007},
          h5py \citep{h5py2021},
          emcee \citep{emcee2013}
          }




\bibliography{sample631}{}
\bibliographystyle{aasjournal}



\end{document}